\documentclass[12pt]{article}
\usepackage{graphicx}
\usepackage{amsmath}
\usepackage{amsfonts}
\usepackage{amssymb}
\begin{document}

\title{A new expression for the production term in the relativistic model of polyatomic gases}
\author{M.C.Carrisi$^1$,  S. Pennisi$^1$, T. Ruggeri$^2$,  T. Arima$^3$ \\
	\\
	1) Department of di Mathematics and Informatics, University of Cagliari, \\
	09124 Cagliari, Italy;
	carrisi@unica.it;  spennisi@unica.it\\
	2) Department of Mathematics and Alma Mater Research Center on Applied \\ Mathematics AM2 
	, University of
	Bologna, 40126 Bologna, Italy;\\
	tommaso.ruggeri@unibo.it\\
3) Department of Engineering for Innovation, National Institute of \\ Technology,  Tomakomai College,
Tomakomai, Hokkaido 059-1275, Japan\\ arima@tomakomai-ct.ac.jp
}
\date{}
\maketitle
 \small {\em \noindent }
\hspace{-2.5 cm}

\abstract{A new expression is here proposed for the production term Q in the Boltzmann Equation. This new expression satisfies  the H-Theorem (non approximated) and, in the transition to Ordinary Thermodynamics gives a result which is independent on the number N of the moments which are used both if the Maxwellian Iteration or the Eckart method are used.}

\section{Introduction}
The balance equations of Relativistic Extended Termodynamics are obtained in \cite{CPTR1} from the Boltzmann equation
\begin{align}\label{0}
p^\alpha \, \partial_\alpha f = Q \, ,
\end{align}
where $f$ is the distribution function and $Q$ the production term. By multiplying it with $\frac{c}{m^{n-1}} \left(1+ \, \frac{\mathcal{I}}{m \, c^2} \right)^n p^{\mu_1} \cdots p^{\mu_n} $ and integrating in $d \, \mathcal{I} \, d \, \vec{P}$, one obtains 
\begin{align}\label{3}
\begin{split}
& \partial_\alpha \, A^{\alpha \alpha_1 \cdots \alpha_n} = I^{\alpha_1 \cdots \alpha_n}
\, \mbox{with} \, n=0 \, , \, N \, \mbox{and} \\ 
& A^{\alpha_1 \alpha_2 \cdots \alpha_{n}}= \frac{c}{m^{n-2}} \, \int_{\Re^3} \int_{0}^{+ \infty } f \, p^{\alpha_1} \, p^{\alpha_2}\, \cdots \, p^{\alpha_{n}} \left(1+ \, \frac{\mathcal{I}}{m \, c^2} \right)^{n-1} \, \varphi (\mathcal{I}) \, d \, \mathcal{I} \, d \, \vec{P} \, , \\
&  I^{\alpha_1 \cdots \alpha_n}= \frac{c}{m^{n-1}} \, \int_{\Re^3} \int_{0}^{+ \infty } Q \, p^{\alpha_1}\, \cdots \, p^{\alpha_n} \left(1+ \, \frac{\mathcal{I}}{m \, c^2} \right)^n \, \varphi (\mathcal{I}) \, d \, \mathcal{I} \, d \, \vec{P} \, .
\end{split}
\end{align}
In particular, $A^{\alpha_1}=V^{\alpha_1}$, $A^{\alpha_1 \alpha_2}= T^{\alpha_1 \alpha_2}$; in the sequel we will consider \eqref{3}$_{2,3}$ also for $n=0$. 
An expression of $Q$ linearized with respect to equilibrium  was proposed in \cite{PR}  (its non linearized expression is introduced in \cite{Penn} ) and reads
\begin{align}\label{0a}
\begin{matrix}
\hspace{- 3,6 cm} \downarrow \\
Q = \frac{U_\alpha p^\alpha}{c^2 \tau} \left(f_E - f -f_E q_\mu p^\mu \frac{1 + \, \frac{\mathcal{I}}{m \, c^2}}{m \, b \, c^2} \right) \, .
\end{matrix}
\end{align}
However, in \cite{DemPen} it was proved that in this way the Maxwellian Iteration to recover Ordinary Thermodynamics gives expressions for the bulk viscosity, the heat conductivity and the shear viscosity which depend on $N$. The reason of this fact is that the balance equation for the triple tensor becomes 
\begin{align*}
\partial_\alpha \, A^{\alpha \beta \gamma } = I^{\beta \gamma}= \frac{U_\alpha}{c^2 \tau} \, \left( A^{\alpha \beta \gamma }_E - A^{\alpha \beta \gamma } \right) - 3 \frac{U_\alpha q_\mu}{\theta_{1,2} m^2 n \, c^6 \, \tau} \,A^{\alpha \beta \gamma \mu }_E \, .
\end{align*}
Besides the last term (which was necessary in order to achieve zero production of mass and of momentum-energy), the other term at the right hand side give $U_\alpha \, \left( A^{\alpha \beta \gamma } - A^{\alpha \beta \gamma }_E \right)$ instead of $T^{\beta \gamma } - T^{\beta \gamma }_E$; this is caused by the presence of the factor $p^\alpha$ indicated by the arrow in \eqref{0a}. So it is necessary to remove this factor; since $Q$ is a scalar, we have to remove also the factor $U_\alpha$ before it. Moreover, since there is a factor $1 + \, \frac{\mathcal{I}}{m \, c^2}$ for every $p^\alpha$, if we eliminate $p^\alpha$, we must also divide everything by $1 + \, \frac{\mathcal{I}}{m \, c^2}$. A similar method has been used in \cite{Corea}, but in the context of Marle approach whit its consequent limitations.
Inspired by these considerations,
we propose now the following expression for the production term $Q$ of the Boltzmann equation: 
\begin{align}\label{1}
\begin{split}
& Q= \frac{1}{\tau \left( 1 + \, \frac{\mathcal{I}}{m \, c^2} \right)} \left( f_E \, - \, f \,  e^{- \, \frac{1}{k_B} \left[m \,  \tilde{\lambda}  + \left(1+ \, \frac{\mathcal{I}}{m \, c^2} \right) p^\mu \,\tilde{\lambda}_\mu   \right] } \right)  \, , 
\end{split}
\end{align}
and $\tilde{\lambda}$, $\tilde{\lambda}_\mu$ unknown functions  to be determined from the conditions
\begin{align}\label{2}
\tilde{\lambda}^{E}=0 \, , \, \tilde{\lambda}_\mu^{E}=0 \, , \,   I=0 \, , \, I^\beta=  0 \,  .
\end{align}
Here \eqref{2}$_{3,4}$ are necessary in order to have zero productions in the conservation laws of mass and of momentum-energy. \\
In the next section we will prove existence and unicity of the solution of \eqref{2} to determine $\tilde{\lambda}$ and $\tilde{\lambda}_\mu$. \\
In section 3 we will prove that the H-Theorem holds up to whatever order with the expression in  \eqref{1}. \\
Now it is well known that the tensors $A^{\alpha \alpha_1 \cdots \alpha_n}$ in the left hand side of eq. \eqref{3}$_1$ can be written as gradients of a 4-potential 
$A^{\alpha \alpha_1 \cdots \alpha_n} = \frac{\partial \, h'^\alpha}{\partial \, \lambda_{\alpha_1 \cdots \alpha_n}}$; a similar thing can be said for the right hand side of eq. \eqref{3}$_1$, if we use the present expression of $Q$. In particular, we will prove in section 4 that the function $Q^*$ exists such that  
 \begin{align}\label{3a}
 I^{\alpha_1 \cdots \alpha_n} = \frac{\partial \, Q^*}{\partial \, \lambda_{\alpha_1 \cdots \alpha_n}}  \, . 
\end{align}
A similar result was found in eq. (5) of \cite{Penn} (or (4)$_2$ with the Landau-Lifshiz approach) starting from the expression of $Q$ in \cite{PR}; we obtain here the same property starting from the present expression of $Q$, with an easier expression. With this same expression we will prove, in section 5 that the Maxwellian Iteration (MI),  when  applied to recover Ordinary Relativistic Thermodynamics, gives expressions for the bulk viscosity $\nu$, the shear viscosity $\mu$ and the heat conductivity which $\chi$  don't depend on $N$. This result is not present with the previous expression of $Q$ proposed in \cite{PR}; in fact, in \cite{DemPen} it has been shown that (MI)  gave an expression for $\chi$ which depended on $N$. In section 6 we will see that, if Ordinary Relativistic Thermodynamics is recovered through the Eckart Method (EK), the  result result is the same of that in (MI). \\
In section 7 the particular case $N=2$ will be considered for a comparison of \eqref{1} with the expression of $Q$ in \cite{PR} and used in \cite{CPTR1}. In section 8 the  non relativistic limit of the field equations, with the new expression \eqref{1} of Q, has been performed. Regarding their left hand sides, we prove that Theorem 1, eq. (12) of \cite{PR1} still holds even if it was found there before that $1+ \, n \, \frac{\mathcal{I}}{m \, c^2}$ was replaced by $\left(1+ \, \frac{\mathcal{I}}{m \, c^2} \right)^n$. What now changes is only the technique proof. So its use in (19) and (20) of \cite{CPTR} is correct. Regarding their right hand sides we  prove here that the limit of $Q$ as expressed by \eqref{1}, for $c \, \rightarrow \, + \infty$ gives  the classical one, as expressed by (50) of \cite{CPTR}. The particular case of a politropic gas is considered in section 9 and, in particular, the case of diatomic gases.
In section 10 the monoatomic limit of the previous results will be calculated and compared with that obtained by working directly with the monoatomic model. In section 11,  the Landau-Liftschitz approach will be considered; it too doesn't depend on $N$.

\section{Existence and unicity of the solution of \eqref{2}}
Let us {\bf firstly} see how \eqref{2} becomes {\bf in the first deviation from equilibrium}. In this case $Q$, proposed in \eqref{1}, becomes
\begin{align}\label{5}
\begin{split}
& Q= \frac{1}{\tau \left( 1 + \, \frac{\mathcal{I}}{m \, c^2} \right)} \left\{  f_E \, - f \, + \, \frac{1}{k_B}  \, f_E \,   \left[m \,  \tilde{\lambda}  + \left(1+ \, \frac{\mathcal{I}}{m \, c^2} \right) p^\mu \,\tilde{\lambda}_\mu   \right]  \right\}  \, , 
\end{split}
\end{align}
It follows that \eqref{2}$_{3,4}$ become
 \begin{align}\label{6}
 \begin{split}
&  A_E \, \tilde{\lambda} \, + \, A^\mu_E \, \tilde{\lambda}_\mu   = \frac{k_B}{m} \, \left( A \, - \, A_E \right) \, , \\
&  V^\beta \, \tilde{\lambda} \, + \, T^{\beta \mu}_E \, \tilde{\lambda}_\mu  = \frac{k_B}{m}  \left( V^\beta \, - \, V^\beta_E \right) = 0 \, .
\end{split}
\end{align}
Moreover, we have 
\begin{align}\label{7}
I^{\beta \gamma} = - \, \frac{1}{m \,  \tau} \left( T^{\beta \gamma} - T^{\beta \gamma}_E \right) +
\frac{1}{k_B \tau} \left( T^{\beta \gamma}_E \, \tilde{\lambda} \, + \, A^{\beta \gamma \mu}_E \, \tilde{\lambda}_\mu  \right) \, .
\end{align}
Here the functions introduced in \eqref{3}$_3$ have been used; the new quantity which is now used is
 \begin{align*}
 \begin{split}
& A_E = m^2 c \, \int_{\Re^3} \int_{0}^{+ \infty } f_E  \left(1+ \, \frac{\mathcal{I}}{m \, c^2} \right)^{-1} \, \varphi (\mathcal{I}) \, d \, \mathcal{I} \, d \, \vec{P} = \rho \, \frac{\int_{0}^{+ \infty} J^*_{2,0} \, \frac{\varphi (\mathcal{I})}{1+ \frac{\mathcal{I}}{m  c^2}} \, d \, \mathcal{I}}{\int_{0}^{+ \infty} J^*_{2,1} \, \varphi (\mathcal{I}) \, d \, \mathcal{I}} \, , \\
& \mbox{with} \,
J_{m,n}(\gamma)= \int_{0}^{+ \infty} \hspace{-0.5 cm} e^{- \gamma \, \cosh \, s} \sinh^m s \cosh^n s \, d \, s \, ,  J^*_{m,n} = J_{m,n}(\gamma^*) \, , \quad \gamma^*= \gamma \left( 1+ \frac{\mathcal{I}}{m  c^2} \right) \, . 
 \end{split}
 \end{align*}
 Moreover, $A \, - \, A_E$ can be desumed with the same passages used  in \cite{CPTR1} to desume $A^{\beta \gamma \mu} - \, A^{\beta \gamma \mu}_E$. Now \eqref{6}$_2$ contracted with $h^\theta_\beta$ gives $h^\theta_\mu \tilde{\lambda}_\mu=0$, so that $\tilde{\lambda}_\mu= \tilde{\lambda}_1 \, \frac{U_\mu}{c^2}$ and the system \eqref{6} becomes 
  \begin{align*}
 \left( \begin{matrix}
  \frac{A_E}{\rho} && 1 \\
  && \\
  1 && \frac{e}{\rho \, c^2}
  \end{matrix} \right)  \left( \begin{matrix}
  \tilde{\lambda}  \\
   \\
  \tilde{\lambda}_1
  \end{matrix} \right) = \frac{k_B}{m} \left( \begin{matrix}
 \frac{A \, - \, A_E}{\rho}  \\
  \\
  0
  \end{matrix} \right) \, ,
  \end{align*}
  from which
 \begin{align}\label{8}
 \tilde{\lambda} = \frac{k_B}{m}  \, \frac{1}{\frac{A_E \, e}{\rho^2  c^2} \, - \, 1}  \, \frac{e}{\rho \, c^2} \, \frac{A \, - \, A_E}{\rho}  \, , \quad  \tilde{\lambda}_1 = - \, \frac{k_B}{m}  \, \frac{1}{\frac{A_E \, e}{\rho^2  c^2} \, - \, 1}  \,  \frac{A \, - \, A_E}{\rho}  \, .
 \end{align}
 By using this result, \eqref{7} becomes
\begin{align}\label{9}
\begin{split}
& I^{\beta \gamma} = - \, \frac{1}{m \,  \tau} \left( T^{\beta \gamma} - T^{\beta \gamma}_E \right) +
 \frac{1}{m \tau \, \rho} 
  \, X^{\beta \gamma}  \left( A \, - \, A_E \right) \quad \mbox{with} \quad \\
& X^{\beta \gamma} =  \frac{1}{\frac{A_E \, e}{\rho^2  c^2} \, - \, 1}  \, \left( T^{\beta \gamma}_E \, \frac{e}{\rho \, c^2}   \, - \, A^{\beta \gamma \mu}_E  \, \frac{U_\mu}{c^2}   \right) \,  \, .
\end{split}
\end{align}
 Besides this linearized expression of $I^{\beta \gamma}$ which will be used later, we have seen that \eqref{2}  gives one and only one solution. \\
{ \bf Let us now apply an iterative procedure } and suppose to have been found $\tilde{\lambda}^{(h)}$ and $\tilde{\lambda}^{(h)}_\mu$, i.e., the expressions of  $\tilde{\lambda}$ and $\tilde{\lambda}_\mu$ up to the order $h$ with respect to equilibrium. By imposing the system \eqref{2} at the order $h+1$ we obtain the system 
 \begin{align}\label{8a}
\begin{split}
&  A_E \, \tilde{\lambda}^{(h+1)} \, + \, A^\mu_E \, \tilde{\lambda}_\mu^{(h+1)}  = \phi_{(h)} \, , \,   A^\beta_E \, \tilde{\lambda}^{(h+1)}  \, + \, A^{\beta \mu}_E \, \tilde{\lambda}_\mu^{(h+1)}     = \phi_{(h)}^\beta \, , 
\end{split}
\end{align}
where the right hand sides are known terms depending also on $\lambda^{*(k)}$, $\lambda^{*(k)}_\mu$  for \\
$k=1, \cdots , h$. Since the matrix of the unknowns $\tilde{\lambda}^{(h+1)}$, $\tilde{\lambda}^{(h+1)}_\mu$ and $\tilde{\lambda}^{(h+1)}_{\mu \nu}$ is the same of that in eq. \eqref{6}, we obtain in this case one and only one solution. This completes the proof of existence and unicity of the solution of \eqref{2}.

\section{The H-Theorem}
We consider the quantity
\begin{align*}
\Sigma  = - \, k_B c  \int_{\Re^3} \int_{0}^{+ \infty }  Q \, \ln \, f  \, \varphi (\mathcal{I}) \, d\lambda_{\alpha_1 \cdots \alpha_n} \, \mathcal{I} \, d \, \vec{P} \, .
\end{align*}
The H-Theorem holds iff $\Sigma \geq 0$. \\
We add to the above quantity $I \, \left(  \tilde{\lambda} \,  - \, \lambda^E \right) + I^\mu \, \left(  \tilde{\lambda}_\mu \,  - \, \lambda^E_\mu \right) = 0$ so that it becomes 
\begin{align*}
\begin{split}
& \Sigma  = - \, k_B c  \int_{\Re^3} \int_{0}^{+ \infty } \frac{f_E}{\tau \left(1+ \frac{\mathcal{I}}{m \, c^2} \right)} \, g(x)  \, \varphi (\mathcal{I}) \, d \, \mathcal{I} \, d \, \vec{P} \, , \quad \mbox{with} \\
& g(x) = (1-x) \, \ln \, x \, , \, x= \frac{f}{f_E} \, e^{- \, \frac{1}{k_B} \left[m \,  \tilde{\lambda}  + \left(1+ \, \frac{\mathcal{I}}{m \, c^2} \right) p^\mu \,\tilde{\lambda}_\mu   \right] } \, .
\end{split}
\end{align*}
Now, $g'' (x) = - \, \frac{1}{x} \,  - \, \frac{1}{x^2} \leq 0$ so that $g'(x)$ is a decreseang function; since $g'(1)=0$, we have that $g'(x)>0$ for $x<1$ and $g'(x)<0$ for $x>1$. Hence in $x=1$ the function $g(x)$ has a maximum; but $g(1)= 0$, so that 
$g(x) \leq 0 $ and $g(x) = 0 $ iff $x=1$, i.e., at equilibrium. It follows that $\Sigma \geq 0$, thus completing the proof.

\section{The gradient form for the production terms}
We know that, if the Maximum Entropy Principle is applied, then the distribution function $f$ has the form 
\begin{align}\label{MEP}
f= e^{-1- \, \frac{1}{k_B T} \, \chi} \quad \mbox{with} \quad \chi = \sum_{n=0}^{N} \frac{1}{m^{n-1}} \, \lambda_{\alpha_1 \cdots \alpha_n} p^{\alpha_1} \dots p^{\alpha_n} \, .
\end{align}
We prove now that, in this hypothesis, the balance equations can be written in the elegant form  
\begin{align}\label{3A}
\boxed{\partial_{\alpha} \, \frac{\partial \, h'^\alpha}{\partial \, \lambda_{\alpha_1 \cdots \alpha_n}} =  \frac{\partial \, Q^*}{\partial \,\lambda_{\alpha_1 \cdots \alpha_n}  } \quad \mbox{for} \quad n \geq 2 \, , } 
\end{align}
with the functions
\begin{align}\label{9A}
\begin{split}
& h'^\alpha = - k_B c \int_{\Re^3} \int_{0}^{+ \infty } f \, p^\alpha \, 
\varphi (\mathcal{I}) \, d \, \mathcal{I} \, d \, \vec{P} \, , \\
& Q^* = \frac{1}{m \, \tau} \, \left( \sum_{n=0}^{N} A_E^{\alpha_1 \cdots \alpha_n} \lambda_{\alpha_1 \cdots \alpha_n} + A_E \, \tilde{\lambda} \, + \, V^\mu \, \tilde{\lambda}_\mu  \right) \,  - \, k_B c \int_{\Re^3} \int_{0}^{+ \infty } Q \, 
\varphi (\mathcal{I}) \, d \, \mathcal{I} \, d \, \vec{P} \, .
\end{split}
\end{align}
We see that both  the left and the right hand side of \eqref{3A} are expressed in the form of a gradient. 
Regarding the left hand side of \eqref{3A} it is a well known result in literature and can be easily tested. For the right hand side we have to prove that
\begin{align}\label{9AA}
I^{\alpha_1 \cdots \alpha_n} =\frac{\partial \, Q^*}{\partial \, \lambda_{\alpha_1 \cdots \alpha_n}} \, . 
\end{align}
To this end we note that the independent variables are $\lambda^E$, $\lambda_\mu^E$, $\lambda_{\alpha_1 \cdots \alpha_n}$ with \\
$n \geq 2$, while $\lambda$, $\lambda_\mu$ are expressed in terms of them by the conditions $V^\alpha - V^\alpha_E=0$, $ \lambda_\alpha^E \lambda_\beta^E \left( T^{\alpha \beta} - T^{\alpha \beta}_E \right)=0$. It follows that $f_E$ doesn't depent on $\lambda_{\alpha_1 \cdots \alpha_n}$ for $ n \geq 2$ and that $\frac{\partial \, f}{\partial \, \lambda_{\alpha_1 \cdots \alpha_n}}= - \frac{f}{k_B} \, \frac{\partial \, \chi}{\partial \, \lambda_{\alpha_1 \cdots \alpha_n}}= - \, \frac{f}{k_B} \left[   m \, \frac{\partial \, \lambda}{\partial \, \lambda_{\alpha_1 \cdots \alpha_n}} + \left( 1 + \, \frac{\mathcal{I}}{m \, c^2}\right) p^\mu \frac{\partial \, \lambda_\mu}{\partial \, \lambda_{\alpha_1 \cdots \alpha_n}} + \right.$ \\
$\left. + \frac{\left( 1 + \, \frac{\mathcal{I}}{m \, c^2}\right)^n}{m^{n-1}} \, p^{\alpha_1} \cdots p^{\alpha_n} \right]$. By taking into account these informations, we have that 
\begin{align*}
\begin{split}
& \frac{\partial \, Q^*}{\partial \, \lambda_{\alpha_1 \cdots \alpha_n}} =
 \frac{1}{m \, \tau} \, \left( A_E \, \frac{\partial \, \left(\lambda +  \tilde{\lambda} \right)}{\partial \, \lambda_{\alpha_1 \cdots \alpha_n}}   \, + \, V^\mu \, \frac{\partial \, \left( \lambda_\mu + \tilde{\lambda}_\mu \right)}{\partial \, \lambda_{\alpha_1 \cdots \alpha_n}} + A_E^{\alpha_1 \cdots \alpha_n}   \right) \,  - \\ 
 & k_B c \int_{\Re^3} \int_{0}^{+ \infty }  \hspace{-0.5 cm}\frac{\partial \, Q}{\partial \, \lambda_{\alpha_1 \cdots \alpha_n}}  \, 
\varphi (\mathcal{I}) \, d \, \mathcal{I} \, d \, \vec{P} =  
\frac{A_E}{m \, \tau} \frac{\partial  \left(\lambda +  \tilde{\lambda} \right)}{\partial \, \lambda_{\alpha_1 \cdots \alpha_n}}   +  \frac{V^\mu}{m \, \tau}  \frac{\partial  \left( \lambda_\mu + \tilde{\lambda}_\mu \right)}{\partial \, \lambda_{\alpha_1 \cdots \alpha_n}} + \frac{1}{m \, \tau}  A_E^{\alpha_1 \cdots \alpha_n}   + \\
& + \frac{k_B c}{\tau} \int_{\Re^3} \int_{0}^{+ \infty }  \frac{1}{ 1 + \, \frac{\mathcal{I}}{m \, c^2}} \frac{\partial \, f}{\partial \, \lambda_{\alpha_1 \cdots \alpha_n}}  \, e^{- \frac{1}{k_B} \left[ m \tilde{\lambda} + \left( 1 + \, \frac{\mathcal{I}}{m \, c^2}\right) p^\mu \tilde{\lambda}_\mu  \right]}
\varphi (\mathcal{I}) \, d \, \mathcal{I} \, d \, \vec{P} - \\
&  \frac{c}{\tau} \int_{\Re^3} \int_{0}^{+ \infty }  \frac{f \,  e^{- \frac{1}{k_B} \left[ m \tilde{\lambda} + \left( 1 + \, \frac{\mathcal{I}}{m \, c^2}\right) p^\mu \tilde{\lambda}_\mu  \right]}}{ 1 + \, \frac{\mathcal{I}}{m \, c^2}} \frac{\partial \,\left[ m \tilde{\lambda} + \left( 1 + \, \frac{\mathcal{I}}{m \, c^2}\right) p^\mu \tilde{\lambda}_\mu  \right]}{\partial \, \lambda_{\alpha_1 \cdots \alpha_n}}  \, 
\varphi (\mathcal{I}) \, d \, \mathcal{I} \, d \, \vec{P} = 
\end{split}
\end{align*}
\begin{align*}
\begin{split}
= & \frac{A_E}{m \, \tau} \frac{\partial  \left(\lambda +  \tilde{\lambda} \right)}{\partial \, \lambda_{\alpha_1 \cdots \alpha_n}}   +  \frac{V^\mu}{m \, \tau}  \frac{\partial  \left( \lambda_\mu + \tilde{\lambda}_\mu \right)}{\partial \, \lambda_{\alpha_1 \cdots \alpha_n}} + \frac{1}{m \, \tau}  A_E^{\alpha_1 \cdots \alpha_n}   -  \frac{c}{\tau} \int_{\Re^3} \int_{0}^{+ \infty }  \frac{1}{ 1 + \, \frac{\mathcal{I}}{m \, c^2}} f  \cdot \\
&  \cdot e^{- \frac{1}{k_B} \left[ m \tilde{\lambda} + \left( 1 + \, \frac{\mathcal{I}}{m \, c^2}\right) p^\mu \tilde{\lambda}_\mu  \right]}
\left[   m \, \frac{\partial \, \lambda}{\partial \, \lambda_{\alpha_1 \cdots \alpha_n}} + \left( 1 + \, \frac{\mathcal{I}}{m \, c^2}\right) p^\mu \frac{\partial \, \lambda_\mu}{\partial \, \lambda_{\alpha_1 \cdots \alpha_n}} + \right. \\
& \left. + \frac{\left( 1 + \, \frac{\mathcal{I}}{m \, c^2}\right)^n}{m^{n-1}} \, p^{\alpha_1} \cdots p^{\alpha_n} \right]
\varphi (\mathcal{I}) 
\, d \, \mathcal{I} \, d \, \vec{P} - \\
&  \frac{c}{\tau} \int_{\Re^3} \int_{0}^{+ \infty }  \frac{f \,  e^{- \frac{1}{k_B} \left[ m \tilde{\lambda} + \left( 1 + \, \frac{\mathcal{I}}{m \, c^2}\right) p^\mu \tilde{\lambda}_\mu  \right]}}{ 1 + \, \frac{\mathcal{I}}{m \, c^2}} \frac{\partial \,\left[ m \tilde{\lambda} + \left( 1 + \, \frac{\mathcal{I}}{m \, c^2}\right) p^\mu \tilde{\lambda}_\mu  \right]}{\partial \, \lambda_{\alpha_1 \cdots \alpha_n}}  \, 
\varphi (\mathcal{I}) \, d \, \mathcal{I} \, d \, \vec{P} = 
\end{split}
\end{align*}
\begin{align*}
\begin{split}
	= & \frac{A_E}{m \, \tau} \frac{\partial  \left(\lambda +  \tilde{\lambda} \right)}{\partial \, \lambda_{\alpha_1 \cdots \alpha_n}}   +  \frac{V^\mu}{m \, \tau}  \frac{\partial  \left( \lambda_\mu + \tilde{\lambda}_\mu \right)}{\partial \, \lambda_{\alpha_1 \cdots \alpha_n}} + \frac{1}{m \, \tau}  A_E^{\alpha_1 \cdots \alpha_n}   -  \frac{c}{\tau} \int_{\Re^3} \int_{0}^{+ \infty }  \frac{1}{ 1 + \, \frac{\mathcal{I}}{m \, c^2}} f  \cdot \\
	&  \cdot e^{- \frac{1}{k_B} \left[ m \tilde{\lambda} + \left( 1 + \, \frac{\mathcal{I}}{m \, c^2}\right) p^\mu \tilde{\lambda}_\mu  \right]}
	\left[   m \, \frac{\partial \, \left( \lambda + \tilde{\lambda} \right)}{\partial \, \lambda_{\alpha_1 \cdots \alpha_n}} + \left( 1 + \, \frac{\mathcal{I}}{m \, c^2}\right) p^\mu \frac{\partial \, \left(\lambda_\mu + \tilde{\lambda}_\mu \right)}{\partial \, \lambda_{\alpha_1 \cdots \alpha_n}} + \right. \\
	& \left. + \frac{\left( 1 + \, \frac{\mathcal{I}}{m \, c^2}\right)^n}{m^{n-1}} \, p^{\alpha_1} \cdots p^{\alpha_n} \right]
	\varphi (\mathcal{I}) 
	\, d \, \mathcal{I} \, d \, \vec{P} \, . 
\end{split}
\end{align*}
Here we insert now 
\begin{align*}
f \,  e^{- \, \frac{1}{k_B} \left[m \,  \tilde{\lambda}  + \left(1+ \, \frac{\mathcal{I}}{m \, c^2} \right) p^\mu \,\tilde{\lambda}_\mu   \right] }  = - \, Q  \, \tau \left( 1 + \, \frac{\mathcal{I}}{m \, c^2} \right) +  f_E  \, , 
\end{align*}
which comes from \eqref{1}, so that it becomes 
\begin{align*}
\begin{split}
& \frac{\partial \, Q^*}{\partial \, \lambda_{\alpha_1 \cdots \alpha_n}} =
 \frac{A_E}{m \, \tau} \frac{\partial  \left(\lambda +  \tilde{\lambda} \right)}{\partial \, \lambda_{\alpha_1 \cdots \alpha_n}}   +  \frac{V^\mu}{m \, \tau}  \frac{\partial  \left( \lambda_\mu + \tilde{\lambda}_\mu \right)}{\partial \, \lambda_{\alpha_1 \cdots \alpha_n}} + \frac{1}{m \, \tau}  A_E^{\alpha_1 \cdots \alpha_n}   +  c \, \int_{\Re^3} \int_{0}^{+ \infty }  Q  \cdot \\
&  \cdot 
\left[   m \, \frac{\partial \, \left( \lambda + \tilde{\lambda} \right)}{\partial \, \lambda_{\alpha_1 \cdots \alpha_n}} + \left( 1 + \, \frac{\mathcal{I}}{m \, c^2}\right) p^\mu \frac{\partial \, \left(\lambda_\mu + \tilde{\lambda}_\mu \right)}{\partial \, \lambda_{\alpha_1 \cdots \alpha_n}} + 
 \frac{\left( 1 + \, \frac{\mathcal{I}}{m \, c^2}\right)^n}{m^{n-1}} \, p^{\alpha_1} \cdots p^{\alpha_n} \right]
\varphi (\mathcal{I}) 
\, d \, \mathcal{I} \, d \, \vec{P} - \\
& \frac{c}{\tau} \int_{\Re^3} \int_{0}^{+ \infty }  \frac{f_E}{ 1 + \, \frac{\mathcal{I}}{m \, c^2}} \cdot \\
&  \cdot 
\left[   m \, \frac{\partial \, \left( \lambda + \tilde{\lambda} \right)}{\partial \, \lambda_{\alpha_1 \cdots \alpha_n}} + \left( 1 + \, \frac{\mathcal{I}}{m \, c^2}\right) p^\mu \frac{\partial \, \left(\lambda_\mu + \tilde{\lambda}_\mu \right)}{\partial \, \lambda_{\alpha_1 \cdots \alpha_n}} + 
\frac{\left( 1 + \, \frac{\mathcal{I}}{m \, c^2}\right)^n}{m^{n-1}} \, p^{\alpha_1} \cdots p^{\alpha_n} \right]
\varphi (\mathcal{I}) 
\, d \, \mathcal{I} \, d \, \vec{P}		\, . 
\end{split}
\end{align*}
Here the last two lines are the opposite of the first three elements on the right hand side of the first line; so there remains 
\begin{align*}
\begin{split}
& \frac{\partial \, Q^*}{\partial \, \lambda_{\alpha_1 \cdots \alpha_n}} =
 c \, \int_{\Re^3} \int_{0}^{+ \infty }  Q  
\left[   m \, \frac{\partial \, \left( \lambda + \tilde{\lambda} \right)}{\partial \, \lambda_{\alpha_1 \cdots \alpha_n}} + \left( 1 + \, \frac{\mathcal{I}}{m \, c^2}\right) p^\mu \frac{\partial \, \left(\lambda_\mu + \tilde{\lambda}_\mu \right)}{\partial \, \lambda_{\alpha_1 \cdots \alpha_n}} + \right. \\
& \left. + \frac{\left( 1 + \, \frac{\mathcal{I}}{m \, c^2}\right)^n}{m^{n-1}} \, p^{\alpha_1} \cdots p^{\alpha_n} \right]
\varphi (\mathcal{I}) 
\, d \, \mathcal{I} \, d \, \vec{P} = 	 I \, \frac{\partial \, \left( \lambda + \tilde{\lambda} \right)}{\partial \, \lambda_{\alpha_1 \cdots \alpha_n}} +I^\mu \frac{\partial \, \left(\lambda_\mu + \tilde{\lambda}_\mu \right)}{\partial \, \lambda_{\alpha_1 \cdots \alpha_n}} + \\
&  +I^{\alpha_1 \cdots \alpha_n} = I^{\alpha_1 \cdots \alpha_n} \, . 
\end{split}
\end{align*}
This completes the proof of \eqref{9AA}.

 \section{Ordinary Relativistic Themodynamics recovered through the Maxwellian Iteration}
 In (MI) the balance equations \eqref{3}$_1$ are written with the left hand sides  calculated at equilibrium and the right hand sides at first order; the terms of the first order with respect to equilibrium are called first iterates and we will denote them with the apex "(OT)". These equations are  
  \begin{align}\label{11}
\partial_\alpha V^\alpha_E = 0 \, , \, \partial_\alpha T^{\alpha \beta}_E = 0 \, , \,   \partial_\alpha A^{\alpha \beta \gamma}_E = I^{(OT) \beta \gamma} \, , \,   \partial_\alpha A^{\alpha \alpha_1 \cdots \alpha_n}_E = I^{(OT) \alpha_1 \cdots \alpha_n}
 \, \mbox{with} \, n=3, \cdots  ,  N .
 \end{align}
 The first 2 of these equations  have the solution 
 \begin{align}\label{2c}
 \begin{split}
 & U^\alpha \, \partial_\alpha \lambda^E =  \left| \begin{matrix}
 \rho && \frac{e}{c^2} \\
 && \\
 \frac{e}{c^2} && \rho \, \theta_{0,2}
 \end{matrix} \right|^{-1} \, \left| \begin{matrix}
 p && \frac{e}{c^2} \\
 && \\
 \frac{1}{3} \, \rho \, c^2 \, \theta_{1,2} && \rho \, \theta_{0,2}
 \end{matrix} \right| \, \frac{1}{T}  \, \partial_\alpha U^\alpha \, , \\
 & U^\alpha U^\beta \, \partial_\alpha \lambda^E_\beta = \left| \begin{matrix}
 \rho && \frac{e}{c^2} \\
 && \\
 \frac{e}{c^2} && \rho \, \theta_{0,2}
 \end{matrix} \right|^{-1} \, \left| \begin{matrix}
 \rho && p \\
 && \\
 \frac{e}{c^2} &&  \frac{1}{3} \, \rho \, c^2 \, \theta_{1,2}
 \end{matrix} \right| \, \frac{1}{T}  \, \partial_\alpha U^\alpha \, \\
&  h^{\alpha \theta} \, \partial_\alpha \lambda^E = - \, \frac{2}{3} \, \frac{\rho}{p} \, c^2 \, \theta_{1,2} \, \left(  \frac{-1}{2T} \,  U^{\alpha} \, \partial_\alpha U^\theta \, - \, \frac{c^2}{2 T^2}  \, h^{\theta  \alpha } \, \partial_\alpha T \right) \, .
 \end{split}
 \end{align}
 (See (15) and (16) of \cite{DemPen} for the calculations and take into account that $\lambda^E_\delta= \frac{U_\delta}{T}$). Since  
 \begin{align*}
 \partial_\alpha A^{\alpha \beta \gamma}_E =  - \, \frac{m}{k_B} \left( A^{\alpha \beta \gamma}_E \, \partial_\alpha \, \lambda \, + \, A^{\alpha \beta \gamma \mu}_E \, \partial_\alpha \, \lambda_\mu \right) \, , 
 \end{align*} 
 these equations imply
\begin{align}\label{2d}
\frac{U_\beta \, U_\gamma}{c^4} \, \partial_\alpha A^{\alpha \beta \gamma}_E =
\frac{D_1}{\left| \begin{matrix}
	\rho && \frac{e}{c^2} \\
	&& \\
	\frac{e}{c^2} && \rho \, \theta_{0,2}
	\end{matrix} \right|} \, \frac{\rho}{p} \, \partial_\alpha \, U^\alpha \, , \, 
g_{\beta \gamma} \, \partial_\alpha A^{\alpha \beta \gamma}_E =
\frac{D_2 }{\left| \begin{matrix}
	\rho && \frac{e}{c^2} \\
	&& \\
	\frac{e}{c^2} && \rho \, \theta_{0,2}
	\end{matrix} \right|}\, \frac{\rho}{p} \, \partial_\alpha \, U^\alpha \, , 
\end{align}
\begin{align*}
h_\beta^\theta \, U_\gamma \, \partial_\alpha A^{\alpha \beta \gamma}_E = \left[\frac{1}{9} \, \frac{\rho^2 c^6}{p} \, \left( \theta_{1,2} \right)^2 -  \frac{\rho \, c^4}{6} \, \theta_{1,3} \right] \, \frac{\rho}{p} \, \left( U^\alpha \partial_\alpha \, U^\theta  \, + \, \frac{c^2}{T} \, h^{\mu \theta}  \partial_\mu \, T \right) \, , 
\end{align*}
\begin{align*}
\begin{split}
& h_\beta^{< \theta} \, h_\gamma^{\psi >} \, \partial_\alpha A^{\alpha \beta \gamma}_E = - \, \frac{2}{3} \, \frac{\rho^2 c^4}{p} \,  \theta_{2,3} h^{\alpha < \theta} \, h^{\psi > \mu} \, \partial_{( \alpha} U_{\mu )} \quad \mbox{with} \quad \\
& D_1=  \left| \begin{matrix}
\rho && \frac{e}{c^2} && p\\
&& && \\
\frac{e}{c^2} && \rho \, \theta_{0,2} && \frac{\rho \, c^2}{3} \, \theta_{1,2} \\
&& && \\
\rho \, \theta_{0,2} &&  \rho \, \theta_{0,3} && \frac{\rho \, c^2}{6} \, \theta_{1,3} 
\end{matrix} \right| \, , \, D_2=  c^2 D_1 - \left| \begin{matrix}
\rho && \frac{e}{c^2} && p\\
&& && \\
\frac{e}{c^2} && \rho \, \theta_{0,2} && \frac{\rho \, c^2}{3} \, \theta_{1,2} \\
&& && \\
\rho \, c^2  \theta_{1,2} &&  \rho \, c^2  \frac{1}{2} \theta_{1,3}  && \rho \, c^4  \frac{5}{3} \,  \theta_{2,3}
\end{matrix} \right|  . 
\end{split}
\end{align*}
From the first two of these equations we have also 
\begin{align}\label{2D}
\frac{g_{\beta \gamma} \, \partial_\alpha A^{\alpha \beta \gamma}_E}{U_\beta \, U_\gamma \, \partial_\alpha A^{\alpha \beta \gamma}_E} = c^4 \frac{D_2}{D_1} \quad \rightarrow \quad  g_{\beta \gamma} \, \partial_\alpha A^{\alpha \beta \gamma}_E = \frac{4 c^4 D_2}{4D_1 - c^6 D_2} \,  U_\beta \, U_\gamma \, \partial_\alpha A^{\alpha < \beta \gamma >}_E \, . 
\end{align}
The equations \ref{11}$_{3,4}$, thanks to eqs. \eqref{9} and \eqref{9AA} can be rewritten as 
\begin{align*}
\partial_\alpha A^{\alpha \beta \gamma}_E =  - \, \frac{1}{m \, \tau} \left( T^{\beta \gamma} - T^{\beta \gamma}_E \right)^{(OT)}  + \, 
\frac{1}{m \tau \, \rho} \, X^{\beta \gamma}
\, \left( A \, - \, A_E \right)^{(OT)} \, ,
 \end{align*} 
\begin{align}\label{12}  
\hspace{- 3,5 cm} \partial_\alpha A^{\alpha \alpha_1 \cdots \alpha_n}_E = \left( \frac{\partial^2 \, Q^*}{\partial \, \lambda_{\alpha_1 \cdots \alpha_n} \partial \, \lambda_{\beta_1 \cdots \beta_n}} \right)_E \, \lambda_{\beta_1 \cdots \beta_n}^{(OT)} 
 \, \mbox{with} \, n=3, \cdots  ,  N \, , 
 \end{align}
 In \eqref{12}$_1$ we can separate the contribute of $\left( T^{\beta \gamma} - T^{\beta \gamma}_E \right)^{(OT)}$ and of $\left( A \, - \, A_E \right)^{(OT)}$ in the following way: Let us contract \eqref{12}$_1$ with $g_{\beta \gamma}$ and use the result to obtain 
 \begin{align}\label{12a} 
\left( A \, - \, A_E \right)^{(OT)} = 
\frac{m \tau \, \rho \, g_{\beta \gamma} \partial_\alpha A^{\alpha \beta \gamma}_E + \,\rho \, g_{\beta \gamma} \left( T^{\beta \gamma} - T^{\beta \gamma}_E \right)^{(OT)}}{X^\mu_\mu}  \, .
 \end{align} 
 After that, let us substitute this result in \eqref{12}$_1$ which in this way becomes 
  \begin{align}\label{12b} 
 \left( g_\delta^\beta \, g_\theta^\gamma \, - \, \frac{X^{\beta \gamma} }{X^\mu_\mu} g_{\delta \theta} \right)  \partial_\alpha A^{\alpha \delta \theta}_E = - \, \frac{1}{m \tau}  \left( g_\delta^\beta \, g_\theta^\gamma \, - \, \frac{X^{\beta \gamma} }{X^\mu_\mu} g_{\delta \theta} \right) \left( T^{\delta \theta} - T^{\delta \theta}_E \right)^{(OT)} \, .
 \end{align} 
Now eq. \eqref{12b} contracted with $h_\beta^{< \theta} \, h_\gamma^{\psi >}$, $h_\beta^{\theta} \, U_\gamma$ gives respectively
 \begin{align}\label{13} 
 \begin{split}
 & h_\beta^{< \theta} \, h_\gamma^{\psi >} \, \partial_\alpha A^{\alpha \beta \gamma}_E =  - \,  \frac{1}{m \, \tau} \,  h_\beta^{< \theta} \, h_\gamma^{\psi >} \, \left( T^{\beta \gamma} - T^{\beta \gamma}_E \right)^{(OT)}  \, , \\
& h_\beta^{\theta} \, U_\gamma \, \partial_\alpha A^{\alpha \beta \gamma}_E =  - h_\beta^{\theta} \, U_\gamma \, \frac{1}{m \, \tau} \left( T^{\beta \gamma} -  \, T^{\beta \gamma}_E \right)^{(OT)} \,  . 
 \end{split}
  \end{align} 
Finally, \eqref{12b} contracted with $U_\beta \, U_\gamma$ gives
 \begin{align}\label{14} 
 \begin{split}
 g_{\delta \theta} \, \left( T^{\delta \theta} - T^{\delta \theta}_E \right)^{(OT)} = m \tau  \left( U_\delta \, U_\theta \frac{X^\mu_\mu}{ U_\beta \, U_\gamma X^{\beta \gamma} } \, - \,  g_{\delta \theta} \right)  \partial_\alpha A^{\alpha \delta \theta}_E  \, .
  \end{split}
 \end{align}
By using \eqref{2D}$_2$, this expression is equivalent to 
 \begin{align}\label{14a} 
\begin{split}
U_\delta U_\theta \, \left( T^{< \delta \theta } - T^{< \delta \theta >}_E \right)^{(OT)} =\frac{ m \tau   U_\delta \, U_\theta \,   \partial_\alpha A^{\alpha < \delta \theta >}_E}{4 \, D_1 - c^6 D_2}  \, \left( \frac{- \, X^\mu_\mu}{\frac{U_\beta \, U_\gamma}{c^2} X^{\beta \gamma}} D_1 + c^6 D_2 \right)\, .
\end{split}
\end{align}
Now \eqref{13} and \eqref{14}, by using \eqref{2d}, give  
 \begin{align*}
  \begin{split}
  & h_\beta^{< \theta} \, h_\gamma^{\psi >} \, \left( T^{\beta \gamma} - T^{\beta \gamma}_E \right)^{(OT)} = \frac{2}{3} \, m \, \tau \, \frac{\rho^2 c^4}{p} \,  \theta_{2,3} h^{\alpha < \theta} \, h^{\psi > \mu} \, \partial_{( \alpha} U_{\mu )}  \, , \\
  & h_\beta^{\theta} \, U_\gamma \,  \left( T^{\beta \gamma} -  T^{\beta \gamma}_E \right)^{(OT)}  = - \, m \, \tau \, \left[\frac{1}{9} \, \frac{\rho^2 c^6}{p} \, \left( \theta_{1,2} \right)^2 -  \frac{\rho \, c^4}{6} \, \theta_{1,3} \right] \, \frac{\rho}{p} \, \frac{c^2}{T} \cdot \\
  & \quad \quad \quad \quad \quad \quad \quad \quad \quad \quad \quad \quad \quad \quad \cdot  \left( \frac{T}{c^2} \, U^\alpha \partial_\alpha \, U^\theta  \, + \, h^{\mu \theta}  \partial_\mu \, T \right) \, , \\
  &   g_{\delta \theta} \, \left( T^{\delta \theta} - T^{\delta \theta}_E \right)^{(OT)} = m \tau  \, \frac{D_1 \, \,  \frac{X^\mu_\mu}{ U_\beta \, U_\gamma X^{\beta \gamma} } \, - D_2 }{\left| \begin{matrix}
  	\rho && \frac{e}{c^2} \\
  	&& \\
  	\frac{e}{c^2} && \rho \, \theta_{0,2}
  	\end{matrix} \right|}   \frac{\rho}{p} \, \partial_\alpha \, U^\alpha  \,  , 
  \end{split}
  \end{align*}
   i.e., 
   \begin{align}\label{15} 
\Pi = - \, \nu \, \partial_\alpha \, U^\alpha \, , \, q^\theta = - \, \chi \, h^{\theta \mu} \, \left( \partial_\mu \, T \, - \, \frac{T}{c^2} \, U^\gamma \, \partial_\gamma \, U_\mu \right) \, , \, t_{< \beta \gamma >}= 2 \, \mu \,  h^\alpha_\beta h^\mu_\gamma \, \partial_{< \alpha} U_{\mu >} \, , 
  \end{align}
with the bulk viscosity $\nu$, the heat conductivity $\chi$ and the shear viscosity $\mu$ given by 
   \begin{align}\label{16} 
\begin{split}
& \nu = \frac{m \tau}{3}  \, \frac{D_1 \, \,  \frac{X^\mu_\mu}{ U_\beta \, U_\gamma X^{\beta \gamma} } \, - D_2 }{\left| \begin{matrix}
	\rho && \frac{e}{c^2} \\
	&& \\
	\frac{e}{c^2} && \rho \, \theta_{0,2}
	\end{matrix} \right|}   \frac{\rho}{p}  \, , \\
& \chi = - \, m \, \tau \, \left[\frac{1}{9} \, \frac{\rho^2 c^6}{p} \, \left( \theta_{1,2} \right)^2 -  \frac{\rho \, c^4}{6} \, \theta_{1,3} \right] \, \frac{\rho}{p} \, \frac{c^2}{T} \, , \quad \mu = \frac{1}{3} \, m \, \tau \, \frac{\rho^2 c^4}{p} \,  \theta_{2,3} \, . 
   \end{split}
\end{align}
So our goal, to find an expression of $Q$ which implies expressions for $\nu$, $\chi$ and  $\mu$ not depending on $N$, has been realized. It is noteworthy that it doesn't suffice to read the coefficients of $\Pi$, $q^\theta$, $t_{< \beta \gamma >}$ from $ \partial_\alpha A^{\alpha \beta \gamma}_E = I^{(OT) \beta \gamma}$; it was necessary, before that, to separate the terms which take a role in $T^{\beta \gamma} - T^{\beta \gamma}_E $. We note also that from eq. (16) of \cite{CPTR1} we have $\frac{e}{\rho c^2}= \omega(\gamma)$, $\theta_{0,2}= \omega^2 - \omega'$ so 
that $- \, \frac{e^2}{\rho \, c^4} \, + \, \rho \, \theta_{0,2}=- \rho \, \omega'$ and \eqref{16}$_1$ is well defined. \\
Finally, \eqref{12}$_2$ and \eqref{12a} are equations to determine $ \lambda_{\beta_1 \cdots \beta_n}^{(OT)} $ which aren't useful to the present goals. \\
We conclude noting that, from \eqref{12a}. \eqref{14} and \eqref{2D}, it follows
\begin{align}\label{16a} 
\frac{\left( A \, - \, A_E \right)^{(OT)}}{\Pi^{(OT)}} = - 3
\frac{\rho \, c^4 \, D_2}{X^\mu_\mu \,\left( \frac{X^\mu_\mu}{ U_\beta \, U_\gamma X^{\beta \gamma} } \, D_1 \, - \, c^4 D_2 \right) } \, - \frac{3 \,\rho}{X^\mu_\mu } \, .
\end{align}

\section{Ordinary Relativistic Themodynamics recovered through the Eckart Method}
This method starts from the conservation laws of mass, of momentum energy  and from the Boltzmann equation, i.e., 
 \begin{align*}
& \partial_\alpha \,V^\alpha = 0 \, , \, \partial_\alpha \,T^{\alpha \beta} = 0 \, , \, p^\alpha \, \partial_\alpha \, f = Q\, .
\end{align*}
At the second step, the left hand sides of these equations are calculated at equilibrium and the right hand sides at first order: 
 \begin{align}\label{10a}
\begin{split}
& \partial_\alpha \,V^\alpha_E = 0 \, , \, \partial_\alpha \,T^{\alpha \beta}_E = 0 \, ,   \\ 
& p^\alpha \, \partial_\alpha \, f_E =  \frac{f_E \, - f }{\tau \left( 1 + \, \frac{\mathcal{I}}{m \, c^2} \right)} + \frac{f_E }{k_B \tau \left( 1 + \, \frac{\mathcal{I}}{m \, c^2} \right)} \left[m \,  \tilde{\lambda}  + \left(1+ \, \frac{\mathcal{I}}{m \, c^2} \right) p^\mu \,\tilde{\lambda}_\mu   \right]   \, ,
\end{split}
\end{align}
from which we desume 
 \begin{align}\label{10b}
f- f_E= - \, \tau \left(1+ \, \frac{\mathcal{I}}{m \, c^2} \right) p^\mu \, \partial_\mu \, f_E \, + \, \frac{f_E}{k_B}  \left[[m \, \tilde{\lambda } + \left(1+ \, \frac{\mathcal{I}}{m \, c^2} \right) p^\mu  \tilde{\lambda }_\mu  \right] \, .
\end{align}
As a test, we multiply this equation times $mc p^\beta \varphi (\mathcal{I})$ and integrate; so we obtain 
\begin{align*}
V^\beta-V^\beta_E = - \, m \, \tau  \, \partial_\mu \,T^{\mu \beta}_E \, + \, 
\frac{m}{k_B}  \left( V^\beta_E \, \tilde{\lambda } + T^{\mu \beta}_E   \tilde{\lambda }_\mu \right) = 0 \, ,
\end{align*}
for eq. \eqref{6}$_2$; this confirms an expected result. \\
We multiply now \eqref{10b} times $c p^\beta p^\gamma \left(1+ \, \frac{\mathcal{I}}{m \, c^2} \right) \varphi (\mathcal{I})$ and integrate; so we obtain 
\begin{align*}
T^{\beta \gamma}-T^{\beta \gamma}_E = - \, m \, \tau  \, \partial_\mu \,A^{\mu \beta \gamma}_E \, + \, 
\frac{m}{k_B}  \left( T^{\beta \gamma}_E \, \tilde{\lambda }  + A^{\mu \beta \gamma}_E   \tilde{\lambda }_\mu \right)  \, .
\end{align*}
Here we can substitute $\tilde{\lambda}_\mu= \tilde{\lambda}_1 \, \frac{U_\mu}{c^2}$ jointly with $\tilde{\lambda}$, $\tilde{\lambda}_\mu$ obtained from \eqref{8} and find the same expression \eqref{12}$_1$ of the Maxwellian Iteration. From now on we have to repeat the passages of the previous section from 
\eqref{12}$_1$ forwards; so it is obvious that the result will be the same, i.e., eqs. \eqref{15} and \eqref{16}. In particular, eq.  \eqref{12}$_1$ contracted with $U_\alpha U_\beta$ gives $\left( A \, - \, A_E \right)^{(OT)}$; this value, substituted again in \eqref{12}$_1$ allows to determine $\left( T^{\beta \gamma} - T^{\beta \gamma}_E \right)^{(OT)}$. \\
It is very satisfactory that the new expression of $Q$ presented here in eq. \eqref{1} give the same result with the Maxwellian Iteration  and with the Eckart approach; moreover, with the Maxwellian Iteration the functions  $\nu$, $\chi$ and  $\mu$ don't depend on $N$ as with previous expression of $Q$ proposed in literature. \\
Note that neither the balance equation $\partial_\alpha \, A^{\alpha \beta \gamma} = I^{\beta \gamma}$, nor  $\partial_\alpha \, A^{\alpha < \beta \gamma >} = I^{< \beta \gamma>}$, nor  $\partial_\alpha \, {A^{\alpha \beta}}_\beta = {I^{\beta}}_\beta$ was taken into account in any of the previous steps. So the results of the Eckart approach are the same $\forall \, N$ and also for the 14 moments model and the 6 moments model, without assuming \eqref{16a}. 

\section{The Maxwellian Iteration for the particular case $N=2$}
In this case we can evaluate $A-A_E$ as $A^{\alpha \beta \gamma} - A^{\alpha \beta \gamma}_E$ in (26) of \cite{CPTR1}:
\begin{align}\label{A}
A-A_E = - \, \frac{m}{k_B} \left[  A_E \left( \lambda - \lambda^E \right) + V^\mu \left( \lambda_\mu - \lambda^E_\mu \right)  + T_E^{\mu \nu} \lambda_{\mu \nu} \right] \, .
\end{align}
Jointly with (28)$_{1-4}$ of \cite{CPTR1}, it gives
\begin{align*}
\left| \begin{matrix}
A_E & \rho & \frac{e}{c^2} & \frac{p}{c^2} &   A - A_E \\
& &  &  &\\
 \theta_{0,0} & \theta_{0,1} & \theta_{0,2} & \frac{1}{3} \, \theta_{1,2} & 0\\
& &  &  &\\
 \theta_{0,1} & \theta_{0,2} & \theta_{0,3} & \frac{1}{6} \, \theta_{1,3} & 0 \\
& &  & & \\
\theta_{0,2} & \theta_{0,3} & \theta_{0,4} & \frac{1}{10} \, \theta_{1,4} &  \frac{\Delta}{4 \, \rho \, c^4} \\
& &  &  & \\
\theta_{1,1} & \frac{1}{3} \, \theta_{1,2} & \frac{1}{6} \, \theta_{1,3} & \frac{5}{9} \, \theta_{2,3} & \frac{\Pi}{\rho \, c^2}  
\end{matrix}\right| =0  \, .
\end{align*}
From this equation we can desume $A-A_E$ as a linear combination of $\Delta$ and $\Pi$; so the above found expression for $\left(A-A_E\right)^{(OT)}$ becomes an equation from which to desume $\Delta^{(OT)}$. But neither is necessary for the above Maxwellian Iteration. So in \cite{CPTR1} $A-A_E$ could be taken as independent variable instead of $\Delta$; this was not done there simply because it did not appear among the quantities  playing a role there.

\subsection{The subsystem with 14 moments} 
The subsystem with 14 moments is obtained by taking only the traceless part of the equation for the triple tensor. \\
The equations \eqref{1}, \eqref{2}, \eqref{5}-\eqref{9} still hold, except that from  \eqref{9} we have to take only the traceless part:
\begin{align}\label{914}
I^{< \beta \gamma> } = - \, \frac{1}{m \,  \tau} \left( T^{< \beta \gamma >} - T^{< \beta \gamma >}_E \right) +
\frac{1}{m \tau \rho} 
X^{< \beta \gamma >} \, \left( A \, - \, A_E \right) \, . 
\end{align}
The equations  \eqref{11}$_{1,2}$ and \eqref{2c} still hold and the equation for the triple tensor must be substituted by 
\begin{align}\label{214}
\partial_\alpha \, A^{\alpha < \beta \gamma> } = I^{< \beta \gamma> } \quad \rightarrow \quad \mbox{(for MI)} \quad \partial_\alpha \, A_E^{\alpha < \beta \gamma> } = I^{< \beta \gamma> } \, ;
\end{align}
by contracting this equation by $\frac{U_\beta U_\gamma}{c^4}$,  $\frac{U_\beta}{c^2} \, h_\gamma^\theta$, $h_\beta^{< \theta} h_\gamma^{ \psi >}$ it becomes equivalent to
\begin{align}\label{314}
\begin{split}
& \frac{U_\beta U_\gamma}{c^4} \, \partial_\alpha \, A_E^{\alpha < \beta \gamma> } =  - \, \frac{3 \, \Pi^{(OT)}}{4 \, m \, c^2 \tau}  \, 
+  \, \frac{1}{m \tau \rho}  \, \frac{ U_\beta U_\gamma}{c^4} \, X^{< \beta \gamma >} \, \left( A \, - \, A_E \right)^{(OT)} \, , \\
&  \frac{U_\beta}{c^2} \, h_\gamma^\theta \,  \partial_\alpha \, A_E^{\alpha < \beta \gamma> } =  - \, \frac{U_\beta}{c^2} \, h_\gamma^\theta \, \frac{1}{m \,  \tau} \left( T^{< \beta \gamma >} - T^{< \beta \gamma >}_E \right)^{(OT)} =  \frac{q^{(OT)\theta}}{m \,  c^2 \tau} \, , \\
& h_\beta^{< \theta} h_\gamma^{ \psi >} \,  \partial_\alpha \, A_E^{\alpha < \beta \gamma> } = - \,  \frac{1}{m \,  \tau} \, t^{(OT) < \theta \psi >}  \, .
\end{split}
\end{align}
The second and the third one of these equations are the same of the general case $N \geq 2$, while  the first one is an equation in the 2 unknowns $\Pi^{(OT)}$ and $\left( A \, - \, A_E \right)^{(OT)}$ but we have no way to calculate them separately. Obviously, since they are scalars and are linear and homogeneous with respect to equilibrium, $\frac{\left( A \, - \, A_E \right)^{(OT)}}{\Pi^{(OT)}}$ depends only on $\rho$ and $\gamma$. This function cannot be calculated from the definition of $A \, - \, A_E$ because there is no reason to think that  $\frac{\left( A \, - \, A_E \right)^{(OT)}}{\Pi^{(OT)}}$ is equal to  $\frac{ A \, - \, A_E }{\Pi}$. In fact, in this way we would get the same problem we had with the old version \eqref{0a} of $Q$, i.e. to read $\mu$, $\chi$ and $\nu$ from $\frac{U_\alpha}{c^2 \tau} \, \left( A^{\alpha \beta \gamma }_E - A^{\alpha \beta \gamma } \right)$ instead of $ T^{\beta \gamma }_E - T^{\beta \gamma } $; this resulted in an Ordinary Thermodynamics depending on $N$. But the following theorem can be proved: \\
{\bf Theorem 1:} "The bulk viscosity $\nu$ of the subsystem with 14 moments is the same of the general case $N \geq 2$ if and only if $\frac{\left( A \, - \, A_E \right)^{(OT)}}{\Pi^{(OT)}}$ has the value given by \eqref{16a}." \\
This condition can be considered a inheritance of the initial system from which it was obtained. If the 14 moments model is considered directly, without considering it a subsystem, then this condition can be assumed as a principle, because there is one and only one Ordinary Thermodynamics so that its coefficients $\nu$, $\mu$ and $\chi$ cannot depend on how many equations we are considering to obtain it through the Maxwellian Iteration.

\subsection{The subsystem with 6 moments} 
The subsystem with 6 moments is obtained by taking only the trace of the equation for the triple tensor. \\
The equations \eqref{1}, \eqref{2}, \eqref{5}-\eqref{9} still hold, except that from  \eqref{9} we have to take only the trace:
\begin{align}\label{96}
I^{\beta \gamma} g_{\beta \gamma} = - \, \frac{1}{m \,  \tau} \left( T^{\beta \gamma} - T^{\beta \gamma}_E \right) g_{\beta \gamma} + \, \frac{1}{m \tau \rho} 
X^\beta_\beta \, \left( A \, - \, A_E \right) \, .
\end{align}
The equations  \eqref{11}$_{1,2}$ and \eqref{2c} still hold and the equation for the triple tensor must be substituted by  $\partial_\alpha \, {A^{\alpha \beta}}_\beta = {I^{\beta}}_\beta  \quad \rightarrow \quad \mbox{(For MI)} \quad \partial_\alpha \, {A^{\alpha \beta}_E}_\beta = {I^{\beta}}_\beta$, i.e., for eq. \eqref{96}:
\begin{align}\label{1-6}
\partial_\alpha \, {A^{\alpha \beta}_E}_\beta = \frac{3 \, \Pi^{(OT)}}{m \,  \tau} +
\frac{1}{m \tau \rho} 
X^\beta_\beta  \, \left( A \, - \, A_E \right)^{(OT)} \, . 
\end{align}
As in the above subsection, this  is an equation in the 2 unknowns $\Pi^{(OT)}$ and $\left( A \, - \, A_E \right)^{(OT)}$ but we have no way to calculate them separately and  there is no reason to think that  $\frac{\left( A \, - \, A_E \right)^{(OT)}}{\Pi^{(OT)}}$ is equal to  $\frac{ A \, - \, A_E }{\Pi}$. In any case, the Theorem 1 of the above subsection still holds, i.e., the bulk viscosity $\nu$ of the model with 14 moments is the same of the general case $N \geq 2$ if and only if $\frac{\left( A \, - \, A_E \right)^{(OT)}}{\Pi^{(OT)}}$ has the value given by \eqref{16a}."

\section{The non relativistic limit}
We want to prove here that the limit of $Q$ as expressed by \eqref{1}, for $c \, \rightarrow \, + \infty$ gives  the classical one, as expressed by (50) of \cite{CPTR}. But, before that, we have to find some preliminary results, as the non relativistic limit of the left hand sides of eqs. \eqref{3}$_1$ and that of the Lagrange multipliers. \\
We have that the limit of \eqref{3} for $c \, \rightarrow \, + \infty$ is still (19) and (20) of \cite{CPTR}, even if they were proved in Theorem 1, eq. (12) of \cite{PR1}, before that $1+ \, n \, \frac{\mathcal{I}}{m \, c^2}$ was replaced by $\left(1+ \, \frac{\mathcal{I}}{m \, c^2} \right)^n$. What now changes is only the technique proof. This is now characterized by two changes of variables. \\
{\bf First change:} We decompose $p^\alpha$ as $p^\alpha= m \, \Gamma (\xi) \left( c \, , \, \vec{\xi} \, \right)$ (Here $\Gamma (\xi)$ is the Lorentz factor) and the change of integration variables, from $\vec{p}$ to $\vec{\xi}$ implies that $d \, \vec{P} = \frac{d \, \vec{p}}{p^0} = \frac{m^2 \Gamma^4}{c} \, d \, \vec{\xi}$. After that, we define 
 \begin{align}\label{20}
 \begin{split}
& \tilde{A}^{i_1 \cdots i_h}_n = c^{h-n-1} \, A^{0 i_1 \cdots i_h \stackrel{n-h}{\overbrace{0 \cdots 0}}} = \\
& = m^4 \, \int_{\Re^3} \int_{0}^{+ \infty } f\, \Gamma^{n+6} \, \xi^{i_1}\, \cdots \, \xi^{i_h} \left(1+ \, \frac{\mathcal{I}}{m \, c^2} \right)^{n} \, \varphi (\mathcal{I}) \, d \, \mathcal{I} \, d \, \vec{\xi} \, . 
\end{split}
\end{align}
The corresponding change of the Lagrange multipliers is 
 \begin{align}\label{21}
\begin{split}
& \tilde{\lambda}_{i_1 \cdots i_h}^n =  \left(  \begin{matrix}
n \\ h
\end{matrix}\right) \, c^{n-h} \, \lambda_{ \stackrel{i_1 \cdots i_h \underbrace{0 \cdots 0}}{\quad \quad n-h}}  \, . 
\end{split}
\end{align}
In fact, if we consider the definition of the 4-potential
\begin{align}\label{22}
  \begin{split}
& h'^\alpha = - \, c \, k_B \int_{\Re^3} \int_{0}^{+ \infty } e^{-1- \frac{\chi}{k_B}}  \, p^\alpha \, \varphi (\mathcal{I}) \, d \, \mathcal{I} \, d \, \vec{P} \quad \mbox{with} \\
& \chi=\sum_{n=0}^{N}\frac{1}{m^{n-1}} \left(1+ \, \frac{\mathcal{I}}{m \, c^2} \right)^n p^{\mu_1} \cdots p^{\mu_n} \lambda_{\mu_1 \cdots \mu_n} \, , 
 \end{split}
\end{align}
 and define $H^\alpha_\beta= \mbox{diag} \, (0,1,1,1)$, $t^\alpha \equiv (1,0,0,0)$ we have 
\begin{align*}
 d \, h'^0 = \sum_{n=0}^{N}  A^{0 \alpha_1 \cdots \alpha_n} d \, \lambda_{\alpha_1 \cdots \alpha_n}= \sum_{n=0}^{N}  A^{0 \alpha_1 \cdots \alpha_n} d \, \lambda_{\beta_1 \cdots \beta_n} \left( H_{\alpha_1}^{\beta_1} + t_{\alpha_1}  t^{\beta_1} \right) \cdots \left( H_{\alpha_n}^{\beta_n} + t_{\alpha_n}  t^{\beta_n} \right) = \\
 = \sum_{n=0}^{N} \sum_{h=0}^{n} \left(  \begin{matrix}
 n \\ h
 \end{matrix}\right)  A^{0 \alpha_1 \cdots \alpha_n} d \, \lambda_{\beta_1 \cdots \beta_n} H_{\alpha_1}^{\beta_1} \cdots H_{\alpha_h}^{\beta_h} t_{\alpha_{h+1}}  t^{\beta^{h+1}} \cdots t_{\alpha_n} t^{\beta_n} = \\
 = \sum_{n=0}^{N} \sum_{h=0}^{n} \left(  \begin{matrix}
 n \\ h
 \end{matrix}\right) 
A^{0 a_1 \cdots a_h \stackrel{n-h}{\overbrace{0 \cdots 0}}} \,  d \, \lambda_{a_1 \cdots a_h \stackrel{n-h}{\underbrace{0 \cdots 0}}}
 =  \sum_{n=0}^{N} \sum_{h=0}^{n} \left(  \begin{matrix}
 n \\ h
 \end{matrix}\right) c^{n+1-h} \tilde{A}_n^{a_1 \cdots a_h} \,  d \, \lambda_{a_1 \cdots a_h \stackrel{n-h}{\underbrace{0 \cdots 0}}} \, , 
\end{align*}
where in the last passage \eqref{20}$_1$ has been used; the result is 
\begin{align*}
\frac{1}{c} \, d \, h'^0 = \sum_{n=0}^{N} \sum_{h=0}^{n} \tilde{A}_n^{a_1 \cdots a_h} \,  d \, \tilde{\lambda}_{a_1 \cdots a_h} \, 
\end{align*}
only if  \eqref{21} holds, so proving it. \\ 
Obviously, by using \eqref{21}, the expression of $\chi$ in \eqref{22} changes; it becomes
\begin{align*}
\begin{split}
& \chi=\sum_{n=0}^{N}\frac{1}{m^{n-1}} \left(1+ \, \frac{\mathcal{I}}{m \, c^2} \right)^n p^{\mu_1} \cdots p^{\mu_n} \lambda_{\nu_1 \cdots \nu_n} \left( H^{\nu_1}_{\mu_1} + t^{\nu_1} t_{\mu_1} \right) \, \cdots \, \left( H^{\nu_n}_{\mu_n} + t^{\nu_n} t_{\mu_n} \right) = \\
& = \sum_{n=0}^{N}  \sum_{h=0}^{n} \frac{1}{m^{n-1}} \left(1+ \, \frac{\mathcal{I}}{m \, c^2} \right)^n p^{\mu_1} \cdots p^{\mu_h} p^{\mu_{h+1}} \cdots p^{\mu_n} \lambda_{\nu_1 \cdots \nu_h \nu_{h+1} \cdots \nu_n}
\left(  \begin{matrix}
n \\h
\end{matrix}\right) \cdot \\
& \cdot H^{\nu_1}_{\mu_1} \cdots H^{\nu_h}_{\mu_h} t^{\nu_{h+1}} t_{\mu_{h+1}} \cdots  t^{\nu_n} t_{\mu_n} = \\
& = \sum_{n=0}^{N}  \sum_{h=0}^{n} \frac{1}{m^{n-1}} \left(1+ \, \frac{\mathcal{I}}{m \, c^2} \right)^n p^{i_1} \cdots p^{i_h}  \lambda_{\stackrel{i_1 \cdots i_h \underbrace{0 \cdots 0}}{\quad \quad n-h}} 
\left(  \begin{matrix}
n \\h
\end{matrix}\right) \left(p^0\right)^{n-h} = \\
& = \sum_{n=0}^{N}  \sum_{h=0}^{n} m \, \Gamma^n \left(1+ \, \frac{\mathcal{I}}{m \, c^2} \right)^n \xi^{i_1} \cdots \xi^{i_h}  \lambda_{\stackrel{i_1 \cdots i_h \underbrace{0 \cdots 0}}{\quad \quad n-h}} 
\left(  \begin{matrix}
n \\h
\end{matrix}\right)  c^{n-h} \quad \rightarrow
\end{split}
\end{align*}
\begin{align}\label{24}
& \chi= m \sum_{h=0}^{N}  \sum_{n=h}^{N}  \Gamma^n  \left(1+ \, \frac{\mathcal{I}}{m \, c^2} \right)^n \xi^{i_1} \cdots \xi^{i_h}  \tilde{\lambda}_{i_1 \cdots i_h}^n 
 \, . 
\end{align}
{\bf Second change:} We consider the following invertible linear combinations of $\tilde{A}^{i_1 \cdots i_h}_n$: 
\begin{align}\label{25}
A^{* i_1 \cdots i_h}_k = \left( 2 \, c^2 \right)^k \underline{\sum_{r=0}^{k} \left(  \begin{matrix}
k \\ r
\end{matrix}\right) (-1)^{k-r}} \, \tilde{A}^{i_1 \cdots i_h}_{h+r} \, .
\end{align}
We note that in the underlined terms we have taken the same coefficients of $(1-1)^k$. By substituting $\tilde{A}^{i_1 \cdots i_h}_{h+r}$ from \eqref{20}, it becomes 
\begin{align*}
\begin{split}
& A^{* i_1 \cdots i_h}_k = \underline{\left( 2 \, c^2 \right)^k \sum_{r=0}^{k} \left(  \begin{matrix}
	k \\ r
	\end{matrix}\right) (-1)^{k-r} }\,
 m^4 \, \int_{\Re^3} \int_{0}^{+ \infty } f \, \Gamma^{h+6} \, \xi^{i_1}\, \cdots \, \xi^{i_h} \underline{\Gamma^{r} \left(1+ \, \frac{\mathcal{I}}{m \, c^2} \right)^{r} }\cdot \\
 & \cdot  \left(1+ \, \frac{\mathcal{I}}{m \, c^2} \right)^{h} \, \varphi (\mathcal{I}) \, d \, \mathcal{I} \, d \, \vec{\xi} \, . 
\end{split}
\end{align*}
Here the underlined terms are equal to
\begin{align*}
\begin{split}
\left( 2 \, c^2 \right)^k \left[ \Gamma \left(1+ \, \frac{\mathcal{I}}{m \, c^2} \right) -1 \right]^k= \left( 2 \, c^2 \right)^k \Gamma^k \left( \frac{ \mathcal{I}}{m \, c^2} + \frac{\frac{\xi^2}{c^2}}{1+ \frac{1}{\Gamma}} \right)^k= \Gamma^k \left( \frac{2 \, \mathcal{I}}{m} + \frac{2 \, \xi^2}{1+ \frac{1}{\Gamma}} \right)^k \, .
\end{split}
\end{align*}
So the above expression becomes 
\begin{align}\label{26}
A^{* i_1 \cdots i_h}_k =
m^4 \, \int_{\Re^3} \int_{0}^{+ \infty } f \, \Gamma^{h+k+6} \, \xi^{i_1}\, \cdots \, \xi^{i_h}   \left(1+ \, \frac{\mathcal{I}}{m \, c^2} \right)^{h} \, \left( \frac{2 \, \mathcal{I}}{m} + \frac{2 \, \xi^2}{1+ \frac{1}{\Gamma}} \right)^k \, \varphi (\mathcal{I}) \, d \, \mathcal{I} \, d \, \vec{\xi} \, . 
\end{align}
Since $\lim_{c \, \rightarrow \, + \infty} m^3 f = f^{classic}$, we have that 
\begin{align*}
\lim_{c \, \rightarrow \, + \infty} A^{* i_1 \cdots i_h}_k =
m \, \int_{\Re^3} \int_{0}^{+ \infty } f^{classic}  \, \xi^{i_1}\, \cdots \, \xi^{i_h}  \left( \frac{2 \, \mathcal{I}}{m} + \frac{2 \, \xi^2}{1+ \frac{1}{\Gamma}} \right)^k \, \varphi (\mathcal{I}) \, d \, \mathcal{I} \, d \, \vec{\xi} \, , 
\end{align*}
as in eq. (20) of \cite{PR1}, even if this article was written  before that $1+ \, n \, \frac{\mathcal{I}}{m \, c^2}$ was replaced by $\left(1+ \, \frac{\mathcal{I}}{m \, c^2} \right)^n$. \\
There remains to prove that the linear combination \eqref{25} is invertible. To this end we prove now that its inverse transformation is 
\begin{align}\label{27}
\tilde{A}^{i_1 \cdots i_h}_{h+k}= \underline{\sum_{r=0}^{k} \left(  \begin{matrix}
	k \\ r
	\end{matrix}\right)}  \, A^{* i_1 \cdots i_h}_r  \left(2 \, c^2 \right)^{-r} \, .
\end{align}
(We note that the underlined terms are the same coefficients of $(1+1)^k$). In fact, the right hand side of \eqref{27}, jointly with \eqref{25}, is 
\begin{align*}
\sum_{r=0}^{k} \left(  \begin{matrix}
k \\ r
\end{matrix}\right) \sum_{s=0}^{r} \left(  \begin{matrix}
r \\ s
\end{matrix}\right) (-1)^{r-s} \tilde{A}^{i_1 \cdots i_h}_{h+s}\, .
\end{align*}
Here we can exchange the order of the two summations and, after that, change the index $r$ to $p$ according to the relation $r=s+p$; so it becomes 
\begin{align*}
\sum_{s=0}^{k} \sum_{p=0}^{k-s} \left(  \begin{matrix}
k \\ s+p
\end{matrix}\right) \left(  \begin{matrix}
s+p \\ s
\end{matrix}\right) (-1)^p \tilde{A}^{i_1 \cdots i_h}_{h+s} =  \sum_{s=0}^{k} \left(  \begin{matrix}
k \\ s
\end{matrix} \right) \left[ \sum_{p=0}^{k-s} \left(  \begin{matrix}
k - s\\ p
\end{matrix}\right) (-1)^p \right]\tilde{A}^{i_1 \cdots i_h}_{h+s}  \, .
\end{align*}
But, if $s<k$, the term in square brackets is $(1-1)^{k-s}=0$; so there remains only the value for $s=k$ which is $\tilde{A}^{i_1 \cdots i_h}_{h+k}$ as in the left hand side of \eqref{27}, thus completing its proof. \\
The corresponding change of the Lagrange multipliers is 
\begin{align}\label{28}
\lambda^{* r}_{a_1 \cdots a_h} = \sum_{p=r}^{N-h}    \left(  \begin{matrix}
p\\ r
\end{matrix}\right)   \left(2 \, c^2 \right)^{-r} \, \tilde{\lambda}_{a_1 \cdots a_h}^{h+p}  = \sum_{q=0}^{N-h-r}    \left(  \begin{matrix}
r+q\\ r
\end{matrix}\right)   \left(2 \, c^2 \right)^{-r} \, \tilde{\lambda}_{a_1 \cdots a_h}^{h+r+q} \, .
\end{align}
In fact,  we have 
\begin{align*}
\frac{1}{c} \, d \, h'^0 = \sum_{n=0}^{N} \sum_{h=0}^{n} \tilde{A}_n^{a_1 \cdots a_h} \,  d \, \tilde{\lambda}_{a_1 \cdots a_h}^n  \stackrel{(1)}{=}  \sum_{n=0}^{N} \sum_{h=0}^{n}  \sum_{r=0}^{n-h} \left(  \begin{matrix}
	n-h \\ r
	\end{matrix}\right)  \, A^{* a_1 \cdots a_h}_r  \left(2 \, c^2 \right)^{-r} \,  d \, \tilde{\lambda}_{a_1 \cdots a_h}^n \stackrel{(2)}{=} \\
= \sum_{h=0}^{N}	\sum_{p=0}^{N-h}   \sum_{r=0}^{p} \left(  \begin{matrix}
	p\\ r
	\end{matrix}\right)  \, A^{* a_1 \cdots a_h}_r  \left(2 \, c^2 \right)^{-r} \,  d \, \tilde{\lambda}_{a_1 \cdots a_h}^{h+p} \stackrel{(3)}{=} \\
	= \sum_{h=0}^{N}	\sum_{r=0}^{N-h} \left[ \sum_{p=r}^{N-h}    \left(  \begin{matrix}
	p\\ r
	\end{matrix}\right)   \left(2 \, c^2 \right)^{-r} \,  d \, \tilde{\lambda}_{a_1 \cdots a_h}^{h+p}  \right]\, A^{* a_1 \cdots a_h}_r = \sum_{h=0}^{N}	\sum_{r=0}^{N-h} \, A^{* a_1 \cdots a_h}_r \, d \, \lambda_{a_1 \cdots a_h}^{*r} \, , 
\end{align*}
where, in the passage denoted with $(1)$ we have used \eqref{27} with $k=n-h$, in that denoted with $(2)$ we have changed the order of the first 2 summations ($ \sum_{n=0}^{N} \sum_{h=0}^{n} \quad \rightarrow \quad \sum_{h=0}^{N}	\sum_{n=h}^{N}$ ) and changed index according to the law $n=h+p$, in the passage denoted with $(3)$ we have changed the order of the second and third summation;  the last passage is correct only if \eqref{28} holds, thus proving it. \\
The inverse of \eqref{28} is 
\begin{align}\label{29}
\tilde{\lambda}_{i_1 \cdots i_h}^{h+s} = \sum_{k=s}^{N-h}    \left(  \begin{matrix}
k\\ s
\end{matrix}\right)  (-1)^{k-s} \left(2 \, c^2 \right)^{k} \, \lambda^{* k}_{i_1 \cdots i_h} \, .
\end{align}
In fact, by substituting it in \eqref{28} we obtain 
\begin{align*}
\lambda^{* r}_{a_1 \cdots a_h} = \sum_{p=r}^{N-h}    \left(  \begin{matrix}
p\\ r
\end{matrix}\right)   \left(2 \, c^2 \right)^{-r} \, \sum_{k=p}^{N-h}    \left(  \begin{matrix}
k\\ p
\end{matrix}\right)  (-1)^{k-p} \left(2 \, c^2 \right)^{k} \, \lambda^{* k}_{a_1 \cdots a_h}   \stackrel{(1)}{=} \\
= \sum_{k=r}^{N-h} \sum_{p=r}^{k}    \left(  \begin{matrix}
k\\ r
\end{matrix}\right) \left(  \begin{matrix}
k-r\\ p-r
\end{matrix}\right)   \,    (-1)^{k-p} \left(2 \, c^2 \right)^{k} \, \lambda^{* k}_{a_1 \cdots a_h}   \stackrel{(2)}{=} \\
= \sum_{k=r}^{N-h} \left(  \begin{matrix}
k\\ r
\end{matrix}\right) \left[\sum_{q=0}^{k-r}     \left(  \begin{matrix}
k-r\\ q
\end{matrix}\right)    \,    (-1)^{k-r-q} \right] \left(2 \, c^2 \right)^{k-r} \, \lambda^{* k}_{a_1 \cdots a_h}   \, ,
\end{align*}
where, in the passage denoted with $(1)$ we have changed the order of the summations
and substituted $ \left(  \begin{matrix}
p\\ r
\end{matrix}\right) \left(  \begin{matrix}
k\\ p
\end{matrix}\right)$ with $\left(  \begin{matrix}
k\\ r
\end{matrix}\right) \left(  \begin{matrix}
k-r\\ p-r
\end{matrix}\right)$ and, in the passage denoted with $(2)$ we have changed index according to the law $p=r+q$. Now, the expression in square brackets for $k>r$ is equal to $(1-1)^{k-r}=0$, so that there remains only the term with $k=r$, i.e., $\lambda^{* r}_{a_1 \cdots a_h}$; so we have obtained an identity so proving that \eqref{29} is the inverse of \eqref{28}. \\
By using it, we see that \eqref{24} becomes
\begin{align*}
\chi= m \sum_{h=0}^{N}  \sum_{n=h}^{N}  \Gamma^n  \left(1+ \, \frac{\mathcal{I}}{m \, c^2} \right)^n \xi^{i_1} \cdots \xi^{i_h}   \sum_{k=n-h}^{N-h}    \left(  \begin{matrix}
k\\ n-h
\end{matrix}\right)  (-1)^{k-n+h} \left(2 \, c^2 \right)^{k} \, \lambda^{* k}_{i_1 \cdots i_h}  \stackrel{(1)}{=} \\
= m \sum_{h=0}^{N}  \sum_{p=0}^{N-h}  \Gamma^{h+p}  \left(1+ \, \frac{\mathcal{I}}{m \, c^2} \right)^{h+p} \xi^{i_1} \cdots \xi^{i_h}   \sum_{k=p}^{N-h}    \left(  \begin{matrix}
k\\ p
\end{matrix}\right)  (-1)^{k-p} \left(2 \, c^2 \right)^{k} \, \lambda^{* k}_{i_1 \cdots i_h}  \stackrel{(2)}{=} 
\\
= m \sum_{h=0}^{N}  \sum_{k=0}^{N-h}  \left[ \sum_{p=0}^{k} \Gamma^{p}  \left(1+ \, \frac{\mathcal{I}}{m \, c^2} \right)^{p}        \left(  \begin{matrix}
k\\ p
\end{matrix}\right)  (-1)^{p} \right] \Gamma^{h}  \left(1+ \, \frac{\mathcal{I}}{m \, c^2} \right)^{h} \left(- \, 2 \, c^2 \right)^{k} \cdot \\ 
\cdot \xi^{i_1} \cdots \xi^{i_h} \lambda^{* k}_{i_1 \cdots i_h}  \, , 
\end{align*}
where in the passage denoted with $(1)$ we have changed index according to the law $n=h+p$, in the passage denoted with $(2)$ we have changed the order of the second and third summation. Now the expression inside the square brackets is equal to 
\begin{align*}
\left[ - \, \Gamma  \left(1+ \, \frac{\mathcal{I}}{m \, c^2} \right)       +1 \right]^{k} = (-1)^{k} \left[ \Gamma  \left(1+ \, \frac{\mathcal{I}}{m \, c^2} \right)       -1 \right]^{k} =  (- \Gamma)^{k} \left[  \left(1+ \, \frac{\mathcal{I}}{m \, c^2} \right)       - \, \frac{1}{\Gamma} \right]^{k} = \\
=  (- \Gamma)^{k} \left[ \frac{\mathcal{I}}{m \, c^2} + \frac{1- \frac{1}{\Gamma^2}}{1+ \frac{1}{\Gamma}}   \right]^{k}= \left( \frac{- \Gamma}{2 \, c^2} \right)^{k} \left[ \frac{2 \, \mathcal{I}}{m} +  \frac{2}{1+ \frac{1}{\Gamma}}  \xi^2 \right]^{k} \, ,
\end{align*}
so that there remains
\begin{align}\label{30}
\chi=  m \sum_{h=0}^{N}  \sum_{k=0}^{N-h}   \Gamma^{h+k}  \left(1+ \, \frac{\mathcal{I}}{m \, c^2} \right)^{h}  \, \xi^{i_1} \cdots \xi^{i_h} \lambda^{* k}_{i_1 \cdots i_h} \left( \frac{2 \, \mathcal{I}}{m} +  \frac{2}{1+ \frac{1}{\Gamma}}  \xi^2 \right)^{k} \, .
\end{align}
It follows that 
\begin{align}\label{31}
\lim_{c \, \rightarrow \, + \infty} \chi=  m \sum_{h=0}^{N}  \sum_{k=0}^{N-h}     , \xi^{i_1} \cdots \xi^{i_h} \lambda^{* k}_{i_1 \cdots i_h} \left( \frac{2 \, \mathcal{I}}{m} +    \xi^2 \right)^{k} \, .
\end{align}
Since $\lambda^{* E k}_{i_1 \cdots i_h} \neq 0$ only for $(h,k)= (0,0), \, (1,0), \, (0,1)$, eq. \eqref{30} calculated at equilibrium gives 
\begin{align}\label{32}
\begin{split}
& \chi^E=  m \left[ \lambda^{*E 0} +  \lambda^{* E 0}_{i} \xi^i \,  \Gamma \,  \left(1+ \, \frac{\mathcal{I}}{m \, c^2} \right)  + \Gamma \, \lambda^{*E1} \left( \frac{2 \, \mathcal{I}}{m} +  \frac{2}{1+ \frac{1}{\Gamma}}  \xi^2 \right)  \right] \quad \rightarrow \\ 
& \lim_{c \, \rightarrow \, + \infty} \chi^E=  m \left[ \lambda^{*E 0} +  \lambda^{* E 0}_{i} \xi^i   + \lambda^{*E1} \left( \frac{2 \, \mathcal{I}}{m} +    \xi^2 \right)  \right] \, .
 \end{split}
\end{align}

\subsection{The non relativistic limit of $\tilde{\lambda}$ and $\tilde{\lambda}_\mu$ }
We have that $\lambda^{*0} - \lambda^{* E 0}$, $\lambda^{*0}_{i} - \lambda^{* E 0}_{i}$, $\lambda^{*1} - \lambda^{* E 1}$ are determined by the conditions $0 = V^\beta - V^\beta_E$, $0 =  U_\beta U_\gamma \left( T^{\beta \gamma} - T^{\beta \gamma}_E \right)$ which are equivalent to $0 = \frac{V^0 - V^0_E}{c}$, $0 = V^i - V^i_E$, $0 =  \frac{U_\beta U_\gamma}{c^2} \left( T^{\beta \gamma} - T^{\beta \gamma}_E \right) - 2 U_\beta \left( V^\beta - V^\beta_E \right)
+ c \left( V^0 - V^0_E \right)$; these relations, by using the decomposition $U^\alpha= m \, \Gamma (v) \left( c \, , \, \vec{v} \, \right)$ become 
 \begin{align*}
\begin{split}
& 0 = m^4 \int_{\Re^3} \int_{0}^{+ \infty} \left( f - f_E \right) \Gamma^5 \varphi ( \mathcal{I}) \, d \, \mathcal{I} \, d \, \vec{\xi}  \, , \, 0 = m^4 \int_{\Re^3} \int_{0}^{+ \infty} \left( f - f_E \right) \xi^j \, \Gamma^5 \varphi ( \mathcal{I}) \, d \, \mathcal{I} \, d \, \vec{\xi}  \, ,  \\
& 0 =  U_\beta U_\gamma \left( T^{\beta \gamma} - T^{\beta \gamma}_E \right)  - 2 U_\beta \left( V^\beta - V^\beta_E \right)
+ c \left( V^0 - V^0_E \right) = \\
& =  \Gamma^2 (v) \left[ \left( T^{00} - c V^0 \right) - \left( T^{00}_E - c V^0_E \right) \right]  + 2 \Gamma (v) v_i  \left[\left( \frac{T^{0i}}{c} -  V^i \right) - \left( \frac{T^{0i}_E}{c} - V^i_E \right) \right] + \\
& + \frac{v_i v_j}{c^2} \Gamma^2 (v)  \left( T^{ij} - T^{ij}_E \right) +  \frac{ V^0 - V^0_E }{c} \left(\Gamma - 1 \right)^2 c^2 \, ,  
\end{split}
\end{align*}
whose non relativistic limit is 
 \begin{align}\label{33}
\begin{split}
& 0 = m \int_{\Re^3} \int_{0}^{+ \infty} \left( f - f_E \right)  \varphi ( \mathcal{I}) \, d \, \mathcal{I} \, d \, \vec{\xi}  \, , \, 0 = m \int_{\Re^3} \int_{0}^{+ \infty} \left( f - f_E \right) \xi^j \, \varphi ( \mathcal{I}) \, d \, \mathcal{I} \, d \, \vec{\xi}  \, ,  \\
& 0 = \lim_{c \, \rightarrow \, + \infty} m^4    \int_{\Re^3} \int_{0}^{+ \infty} \left( f - f_E \right)  \left( \frac{\xi^2}{1 + \frac{1}{\Gamma}}+  \frac{ \mathcal{I}}{m}  \right) \Gamma^6 \, \varphi ( \mathcal{I}) \, d \, \mathcal{I} \, d \, \vec{\xi} = \\
& \quad \quad \quad = m \int_{\Re^3} \int_{0}^{+ \infty} \left( f - f_E \right)  \left( \frac{\xi^2}{2}+  \frac{ \mathcal{I}}{m}  \right)  \, \varphi ( \mathcal{I}) \, d \, \mathcal{I} \, d \, \vec{\xi} \, . 
\end{split}
\end{align}
Moreover, $\tilde{\lambda}$ and $\tilde{\lambda}_\mu$ are determined by \eqref{2},  which are equivalent to $I=0$, $I^j=0$, $2 c \, I^0 - 2 c^2  I=0$, i.e., 
 \begin{align}\label{34}
\begin{split}
& 0 = \frac{m^3}{\tau} \int_{\Re^3} \int_{0}^{+ \infty} \frac{\Gamma^4}{1+ \, \frac{\mathcal{I}}{m \, c^2}  }  
\left( f_E  \, - \, f \, e^{- \, \frac{1}{k_B} \left[m \, \tilde{\lambda} + \left(1+ \, \frac{\mathcal{I}}{m \, c^2} \right) p^\mu \tilde{\lambda}_\mu \right] } \right) \varphi ( \mathcal{I}) \, d \, \mathcal{I} \, d \, \vec{\xi}  \, , \\ 
& 0 =  \frac{m^3}{\tau} \int_{\Re^3} \int_{0}^{+ \infty} \, 
\Gamma^5 \left( f_E \, - \, f \, e^{- \, \frac{1}{k_B} \left[m \, \tilde{\lambda} + \left(1+ \, \frac{\mathcal{I}}{m \, c^2} \right) p^\mu \tilde{\lambda}_\mu  \right] } \right) \xi^j \,  \varphi ( \mathcal{I}) \, d \, \mathcal{I} \, d \, \vec{\xi}   \, ,  \\
& 0 = \frac{m^3}{\tau} \int_{\Re^3} \int_{0}^{+ \infty} \frac{\Gamma^5}{1+ \, \frac{\mathcal{I}}{m \, c^2}} \left( \frac{2 \, \xi^2}{1 + \frac{1}{\Gamma}}+  \frac{2 \, \mathcal{I}}{m}  \right)
 \left( f_E \, - \, f \, e^{- \, \frac{1}{k_B} \left[m \, \tilde{\lambda} + \left(1+ \, \frac{\mathcal{I}}{m \, c^2} \right) p^\mu \tilde{\lambda}_\mu  \right] } \right) \cdot \\
 & \quad \quad \quad  \cdot \varphi ( \mathcal{I}) \, d \, \mathcal{I} \, d \, \vec{\xi}   \, ,
\end{split}
\end{align}
It follows that  a  sufficient condition for eqs. \eqref{34} to be satisfied in the non relativistic limit, thanks also to \eqref{33}, is that
\begin{align}\label{35}
\begin{split}
& \lim_{c \, \rightarrow \, + \infty}  \frac{\tilde{\lambda}_0}{c}  =0 \, , \, 
 \lim_{c \, \rightarrow \, + \infty}\left[ \tilde{\lambda} +  c \, \tilde{\lambda}_0  \right] = 0 \, , \,
\lim_{c \, \rightarrow \, + \infty}  \tilde{\lambda}_i =0 \quad \rightarrow \\
& \lim_{c \, \rightarrow \, + \infty}\left[ m \, \tilde{\lambda} + \left(1+ \, \frac{\mathcal{I}}{m \, c^2} \right) p^\mu \tilde{\lambda}_\mu \right] =0 \, .   
\end{split}
\end{align}
Since we have already proved that there is only one solution of the problem, this unique solution is \eqref{35}. It follows also that the non relativistic limit of the definition of $Q$ in \eqref{1} is 
\begin{align}\label{1a}
 \lim_{c \, \rightarrow \, + \infty} Q= \frac{f_E \, - \, f}{\tau }  \, , 
\end{align}
as in eq. (50) of \cite{CPTR}. 

\section{The politropic case}
In this case the measure $\varphi (\mathcal{I})$ is equal to $\mathcal{I}^a$, where $a$ measures how much the gas is poliatomic; in particular, $a=0$ for diatomic gases and monoatomic gases are obtained by taking the limit for $a$ going to $-1$. Since everything is determined in terms of $\omega(\gamma)= \frac{e}{\rho \, c^2}$, it will suffice to see what $\omega(\gamma)$ becomes with $\varphi (\mathcal{I})=\mathcal{I}^a$. We reach this end through the following steps.
\begin{itemize}
	\item Let us consider eq. (7.6)$_2$ and two lines after eq. (7.12) of \cite{MUr}; for $m=0$, $n=2$, it becomes 
\end{itemize}
\begin{align}\label{2a}
\gamma \, J_{2,2} (\gamma) = 3 \, J_{0,3}(\gamma) \, - 2 \, J_{0,1} (\gamma) \quad \rightarrow \quad \gamma \, \left(1+ \, \frac{\mathcal{I}}{m \, c^2} \right) J_{2,2} (\gamma^*) = 3 \, J_{0,3}(\gamma^*) \, - 2 \, J_{0,1^*} (\gamma) \, .
\end{align}
\begin{itemize}
	\item We evaluate $\int_{0}^{+ \infty} J_{0,m}(\gamma^*) \, \mathcal{I}^a d \, \mathcal{I}$ by changing the order of integration and by using the change of integration variables from $\mathcal{I}$ to $x$ according to the law  $\mathcal{I}= \frac{k_B T}{\cosh \, s} \, x$,  as in  \cite{Monolim}: 
\begin{align*}	
\int_{0}^{+ \infty} J_{0,m}(\gamma^*)  \, \mathcal{I}^a d \, \mathcal{I} = 	\int_{0}^{+ \infty} e^{- \gamma \cosh \, s} \left( \int_{0}^{+ \infty} e^{- \frac{\mathcal{I}}{k_B T}{\cosh \, s}} \cosh^m \, s \, \mathcal{I}^a d \, \mathcal{I} \right)  d \, s = \\
= \int_{0}^{+ \infty} e^{- \gamma \cosh \, s} \cosh^{m-a-1} \, s \, d \, s \left( \int_{0}^{+ \infty} e^{- x} x^a d \, x \right) = \\
= \left( k_B T \right)^{a+1} \Gamma (a+1) \int_{0}^{+ \infty} \hspace{- 0.5 cm} e^{- \gamma \cosh \, s} \cosh^{m-a-1}  s \, d \, s = \left( k_B T \right)^{a+1} \Gamma (a+1)  J_{0,m-a-1}(\gamma)  ,
\end{align*}
where the Gamma Function has been used for $\Gamma (a+1)$. \\
By using this result in the previous step, we obtain
\begin{align*}
\gamma \int_{0}^{+ \infty} J_{2,2}(\gamma^*) \,  \left(1+ \, \frac{\mathcal{I}}{m \, c^2} \right) \mathcal{I}^a d \, \mathcal{I} = \left( k_B T \right)^{a+1} \Gamma (a+1)  \left[ 3 \, J_{0,2-a}(\gamma) \, - \, 2 \, J_{0,-a}(\gamma) \right] \, . 
\end{align*}
\item  We evaluate $\int_{0}^{+ \infty} J_{2,1}(\gamma^*) \, \mathcal{I}^a d \, \mathcal{I}$ with  the same passages of the previous step. We find 
\begin{align*}
\int_{0}^{+ \infty} J_{2,1}(\gamma^*) \, \mathcal{I}^a d \, \mathcal{I} = \left( k_B T \right)^{a+1} \Gamma (a+1) \, J_{2,-a}(\gamma)  \, . 
\end{align*}
\end{itemize}
As consequence of these steps, we have 
\begin{align}\label{4a}
\omega(\gamma) = \frac{\int_{0}^{+ \infty} J_{2,2}(\gamma^*) \,  \left(1+ \, \frac{\mathcal{I}}{m \, c^2} \right) \mathcal{I}^a d \, \mathcal{I} }{\int_{0}^{+ \infty} J_{2,1}(\gamma^*) \, \mathcal{I}^a d \, \mathcal{I}}  = \frac{1}{\gamma} \, \frac{3 \, J_{0,2-a}(\gamma) \, - \, 2 \, J_{0,-a}(\gamma) }{ J_{2,-a}(\gamma)} \, . 
\end{align}
It follows that 
\begin{align*}
\lim_{a \, \rightarrow \, -1 }\omega(\gamma) =  \frac{1}{\gamma} \, \frac{3 \, J_{0,3}(\gamma) \, - \, 2 \, J_{0,1}(\gamma) }{ J_{2,1}(\gamma)} =  \frac{ J_{2,2}(\gamma)  }{ J_{2,1}(\gamma)} \, , 
\end{align*}
where in the last passage we have used \eqref{2a}$_1$; the result is what we expected for the monoatomic limit. \\
{\bf It follows also,  for $a=0$ (diatomic gases), that}
\begin{align*}
\omega(\gamma) =  \frac{1}{\gamma} \, \frac{3 \, J_{0,2}(\gamma) \, - \, 2 \, J_{0,0}(\gamma) }{ J_{2,0}(\gamma)} =  \frac{1}{\gamma} \frac{3 \, J_{2,0}(\gamma) \, +  \, J_{0,0}(\gamma) }{ J_{2,0}(\gamma)} = \frac{3}{\gamma} \, + \, \frac{J_{0,0}(\gamma) }{ J_{0,1}(\gamma)}  \, , 
\end{align*}
where we have considered $J_{0,2}= J_{2,0} + J_{0,0}$ which comes from (7.6)$_1$  of \cite{MUr}, for $m=0$, $n=0$ and, in the last passage,
again eq. (7.6)$_2$ and two lines after eq. (7.12) of \cite{MUr}, for $m=0$, $n=0$.
This result is the same of \cite{CPTR1} (written two lines before its eq. (59)) where it was taken from \cite{RXZ}. In fact $J_{0,0}=K_0$, $J_{0,1}=K_1$, where $K_n$
is the modified Bessel function of second kind $K_n(\gamma) = \int_{0}^{+ \infty} \cosh \, (n \, s) e^{- \gamma \, \cosh \, s} d \, s$. We can now call $\frac{J_{0,0}(\gamma) }{ J_{0,1}(\gamma)} = G^*$; by using the calculations of Appendix A, we obtain
\begin{align*}
\begin{split}
& \theta_{0,0}= 1 \, , \, \theta_{0,1}= \omega = \frac{3}{\gamma} \, + \, G^* \, , \, \theta_{0,2}=  \frac{12}{\gamma^2} \, + \, 1  + 5 \, \frac{G^*}{\gamma} \, , \\ 
& \theta_{0,3}=  \left(G^* \right)^2 + 3 G^* \left( 1 + \frac{7}{\gamma^2} \right) + \frac{60}{\gamma^3} + \frac{8}{\gamma}
\, , \, \theta_{1,1}= \frac{1}{\gamma} \, , \,
 \theta_{1,2}= \frac{12}{\gamma^2} + \frac{3 G^*}{\gamma} \, , \\ 
& \theta_{1,3}=   \frac{42}{\gamma^2} \, G^* + \frac{102}{\gamma^3} \, + \frac{18}{\gamma^2} \, + \frac{6}{\gamma} \, , \,
 \theta_{2,3}= \frac{3G^*}{\gamma^2}  +  \frac{12}{\gamma^3} \, , \, p = \frac{\rho \, c^2}{\gamma} \, . 
\end{split}
\end{align*}
By substituting these values in \eqref{16}, we obtain 
\begin{align}\label{5a} 
\begin{split}
& \mu = m \, \tau \, \frac{\rho^2 c^4}{p} \,  \left( \frac{G^*}{\gamma^2}  +  \frac{4}{\gamma^3} \right) \, , \\
& \chi = - \, m \, \tau \, \left[  \frac{\left(G^*\right)^2}{\gamma} + \frac{G^*}{\gamma^2} \,  \left(8 -   \frac{7}{\gamma^2}  \right) - \frac{1}{\gamma^3} \, - \frac{3}{\gamma^2} \, - \frac{1}{\gamma}   \right] \, \frac{\rho^2}{p} \, \frac{c^6}{T} \, , \\
& \nu = m \, \tau \, 
\left\{ \frac{- \, \frac{3}{\gamma^2} }{ \left[ - \left( G^* \right)^2 +   \frac{3}{\gamma^2} \, - \, \frac{G^*}{\gamma}  \right]^2 } \left[ - \, \frac{2}{\gamma^2} \left(G^*\right)^3 +  \left(G^*\right)^2 \left( - \, \frac{7}{\gamma^3} -  \frac{3}{\gamma^2}  \,+ \frac{5}{\gamma} -1 \right) + \right. \right. \\
& \quad \quad \left. + G^* \left(  \frac{1}{\gamma^4} -  \frac{3}{\gamma^3}  \,-  \frac{2}{\gamma^2} \right) - \frac{60}{\gamma^5} +  \frac{36}{\gamma^4}  \,+  \frac{3}{\gamma^2}  \right]  \, +  \\
&  \quad \quad \quad +  \frac{1}{ - \left( G^* \right)^2 +   \frac{3}{\gamma^2} \, - \, \frac{G^*}{\gamma}   } 
\left[  - \, \frac{4}{\gamma^2} \left(G^*\right)^3 -  \frac{11}{\gamma^3}   \left(G^*\right)^2 + 
G^* \left( \frac{4}{\gamma^4} \, + \, \frac{4}{\gamma^2} \, - \, \frac{6}{\gamma^3}\right) \right. + \\
& \quad \quad \left. \left.  + 
  \, \frac{24}{\gamma^5} -  \frac{15}{\gamma^4}  \, - \, \frac{1}{\gamma^3} \, - \, \frac{3}{\gamma^2}  \right] \right\}   \frac{\gamma\, c^2 }{\rho } \, . 
\end{split}
\end{align}
\section{The monoatomic case}
Before doing this limit, let us consider the results obtained by working directly with the monoatomic model. In this case we still have  expressions which can be formally obtained from \eqref{3}$_{2,3}$ by substituting in them $\varphi (\mathcal{I})=1$, after that, $\mathcal{I}=0$ and finally by eliminating the integrations in $d \, \mathcal{I}$. In this way the trace conditions arise  
\begin{align}\label{37}
\begin{split}
& A= \frac{1}{c^2} \, g_{\beta \gamma} T^{\beta \gamma} \, , \, V^\alpha = \frac{1}{c^2} \, g_{\beta \gamma} A^{\alpha \beta \gamma} \, , \, 
 T^{\alpha \beta}  = \frac{1}{c^2} \, g_{\mu \nu} A^{\alpha \beta \mu \nu} \, , \, 
 I^{\alpha \beta}  g_{\alpha \beta} =0 \, , \\
 & X^\mu_\mu = \rho \, c^2 \, ,
 \end{split}
\end{align}
where, for the last one eq. \eqref{9}$_2$ has been used and \eqref{37}$_4$ is automatically satisfied as consequence of \eqref{9}$_1$ and \eqref{37}$_5$. 
Taking into account of this fact, the first three of the balance equations with the left hand sides calculated at equilibrium and the right hand sides at first order with respect to equilibrium can be substituted by 
 \begin{align}\label{38}
 \begin{split}
& \partial_\alpha \, V^{\alpha}_E = 0 \, , \, \partial_\alpha \, T^{\alpha \beta}_E = 0 \, , \\ 
& \partial_\alpha \, A^{\alpha < \beta \gamma >} = I^{< \beta \gamma >} = - \, \frac{1}{m \,  \tau} \left( T^{< \beta \gamma >} - T^{< \beta \gamma >}_E \right) - \frac{3 \Pi}{m c^2 \tau \, \rho} 
\, X^{< \beta \gamma >}  \, . 
 \end{split}
\end{align}
We note that \eqref{2d}$_2$ implies $D_2=0$, so that the condition \eqref{16a} is satisfied and the {\bf Maxwellian Iteration} fits with the general case. \\
Let us now see what happens with the {\bf Eckart Approach}. In this case we start from the equations 
 \begin{align}\label{39a}
\begin{split}
& \partial_\alpha \,V^\alpha_E = 0 \, , \, \partial_\alpha \,T^{\alpha \beta}_E = 0 \, ,  \, p^\alpha \, \partial_\alpha \, f_E = Q \, , 
\end{split}
\end{align}
where $Q$ is given by \eqref{5}. By substituting there $\tilde{\lambda}_\mu= \tilde{\lambda}_1 \, \frac{U_\mu}{c^2}$ and $\tilde{\lambda}$, $\tilde{\lambda}_1$ given by \eqref{8}, it becomes 
\begin{align}\label{39ab}
\begin{split}
& Q= \frac{1}{\tau \left( 1 + \, \frac{\mathcal{I}}{m \, c^2} \right)} \left\{  f_E \, - f \, + \, \frac{1}{m \, \rho}  \,  f_E \,   \left[m  \, \frac{e}{\rho \, c^2} \, - \left(1+ \, \frac{\mathcal{I}}{m \, c^2} \right) p^\mu  \, \frac{U_\mu}{c^2} \,   \right] \frac{A \, - \, A_E}{\frac{A_E \, e}{\rho^2  c^2} \, - \, 1}  \right\}  \, , 
\end{split}
\end{align}
which, for monoatomic gases is 
\begin{align}\label{39ac}
\begin{split}
& Q= \frac{1}{\tau} \left\{  f_E \, - f \, + \, \frac{3}{m \, c^2 \rho}  \,  f_E \,   \left[- m  \, \frac{e}{\rho \, c^2} \, +  p^\mu  \, \frac{U_\mu}{c^2} \,   \right] \frac{\Pi}{\frac{A_E \, e}{\rho^2  c^2} \, - \, 1}  \right\}  \, , 
\end{split}
\end{align}
where we have taken into account of $A-A_E = - \, \frac{3 \, \Pi}{c^2}$ which comes from \eqref{37}$_1$. \\
So, from \eqref{39a}$_3$ we desume
\begin{align}\label{39ad}
\begin{split}
& f \, - f_E  = - \, \tau \, p^\mu \, \partial_\mu \, f_E  \, + \, \frac{3}{m \, c^2 \rho}  \,  f_E \,   \left[- m  \, \frac{e}{\rho \, c^2} \, +  p^\mu  \, \frac{U_\mu}{c^2} \,   \right] \frac{\Pi}{\frac{A_E \, e}{\rho^2  c^2} \, - \, 1}    \, . 
\end{split}
\end{align}
By multiplying this equation times $m \, c \, p^\alpha$ and integrating in $d \, \vec{P}$ we obtain 
\begin{align*}
0 = V^\alpha - V^\alpha_E = - \, m \, \tau \, \partial_\mu \, T^{\mu \alpha}_E \, + \, \frac{3}{c^2 \rho}   \left[-  \, \frac{e}{\rho \, c^2} \, V^\alpha_E + \, T^{\mu \alpha}_E \, \frac{U_\mu}{c^2} \,   \right] \frac{\Pi}{\frac{A_E \, e}{\rho^2  c^2} \, - \, 1} \, ,
\end{align*}
which is an identity, thanks also to \eqref{39a}$_2$. \\
By multiplying \eqref{39ad} times $c \, p^\alpha p^\beta$ and integrating in $d \, \vec{P}$ we obtain 
\begin{align}\label{39ae}
\begin{split}
& T^{\alpha \beta} \, - T^{\alpha \beta}_E  = - m \, \, \tau  \, \partial_\mu \, A^{\mu \alpha \beta}_E \, + \, \frac{3}{c^2 \rho}    \left[-  \, \frac{e}{\rho \, c^2} \, T^{\alpha \beta}_E  +  A^{\mu \alpha \beta}_E   \, \frac{U_\mu}{c^2} \,   \right] \frac{\Pi}{\frac{A_E \, e}{\rho^2  c^2} \, - \, 1}    \, . 
\end{split}
\end{align}
The trace of this relation is an identity, thanks to \eqref{37}$_2$, \eqref{39a}$_1$ and \eqref{37}$_1$ calculated at equilibrium; so it is equivalent to its traceless part, i.e., 
\begin{align}\label{39af}
\begin{split}
& T^{< \alpha \beta >} \, - T^{< \alpha \beta >}_E  = - m \, \tau  \, \partial_\mu \, A^{\mu < \alpha \beta >}_E \, + \, \frac{3}{c^2 \rho}    \left[-  \, \frac{e}{\rho \, c^2} \, T^{< \alpha \beta >}_E  +  A^{\mu < \alpha \beta >}_E   \, \frac{U_\mu}{c^2} \,   \right] \frac{\Pi}{\frac{A_E \, e}{\rho^2  c^2} \, - \, 1}    \, . 
\end{split}
\end{align}
This is the same result \eqref{38}$_3$ of  the Maxwellian Iteration, if we take into account the expression \eqref{9}$_2$ of $X^{\beta \gamma}$.

\section{The Landau-Lifshiz description} 
From articles such as \cite{CarIo} and \cite{Penn} we know that with this desciption nothing changes with respect to the above used model, if the Lagrange multipliers are maintained as independent variables. Also the variables at equilibrium are the same; the difference comes out when the deviations of the variables from equilibrium are defined. More precisely, both approaches use as independent variables $\rho$, $U^\alpha$, $\gamma$, $\lambda - \lambda^E$, $\lambda_\mu - \lambda^E_\mu$, $\lambda_{\mu_1 \cdots \mu_n} $ with $n=2 \, , \, \cdots \, , \, N$ which are constrained according to the following scheme \\

\begin{tabular}{|c|c|}
	\hline 
	& \\
	Constraints with Eckart approach	&  Constraints with Landau-Lifshiz approach \\ 
	& \\
	\hline 
	& \\
	$V^\alpha - V^\alpha_E=0$	&  $U_\alpha \left( V^\alpha - V^\alpha_E \right)=0$ \\
	& \\
	$U_\alpha U_\beta \left( T^{\alpha \beta} - T^{\alpha \beta}_E \right)=0$ &  $U_\beta \left( T^{\alpha \beta} - T^{\alpha \beta}_E \right)=0$ \\
	\hline 
\end{tabular} 
\\
\\
\\
In other words, the constraints    $U_\alpha \left( V^\alpha - V^\alpha_E \right)=0$,  $U_\alpha U_\beta \left( T^{\alpha \beta} - T^{\alpha \beta}_E \right)=0$ are present in both approaches, while the constraint $h_\alpha^\theta \left( V^\alpha - V^\alpha_E\right)=0$ of the Eckart approach is substituted in the Landau-Lifshiz approach with the constraint $h_\alpha^\theta U_\beta \left( T^{\alpha \beta} - T^{\alpha \beta}_E \right)=0$. After that, the deviation from equilibrium is defined as
\begin{align*}
V^\alpha - V^\alpha_E = q^\alpha \quad , \quad  T^{\alpha \beta} - T^{\alpha \beta}_E= \Pi \,  h^{\alpha \beta} \, + \, t^{< \alpha \beta >_3} \, . 
\end{align*}
Obviously, these $\Pi$, $q^\alpha$, $t^{< \alpha \beta >_3} $ are different from thosee of the Eckart approach even if, as shown in \cite{Penn}, there is an invertible transformation law between them; here we have called them in the same way in order not to have an heavy notation. \\
It is evident that the conditions \eqref{2}, \eqref{6}$_1$, \eqref{6}$_2$ contracted with $U_\beta$  and the expression \eqref{7} still hold, while \eqref{6}$_2$ contracted with $h_\beta^\theta$ has to be replaced by
\begin{align}\label{50}
h^{\beta \theta} \, \tilde{\lambda}_\mu = \frac{k_B}{m \, p} \, q^\theta \, . 
\end{align} 
Therefore we have $\tilde{\lambda}_\mu = - \, \frac{k_B}{m \, p} \, q_\mu + \tilde{\lambda}_1 \frac{U_\mu}{c^2}$ with $\tilde{\lambda}$ and $\tilde{\lambda}_1$ still given by \eqref{8}. In this way \eqref{7} becomes 
\begin{align}\label{9a}
\begin{split}
& I^{\beta \gamma} = - \, \frac{1}{m \,  \tau} \left( T^{\beta \gamma} - T^{\beta \gamma}_E \right) + \, \frac{2}{3 \, k_B \tau} \, \rho \,  c^2 \, \theta_{1,2} \, U^{( \beta} h^{\gamma ) \mu} \tilde{\lambda}_\mu + \\
& \quad \quad \quad + \, \frac{1}{k_B \tau} \left[T_E^{\beta \gamma} \, \tilde{\lambda} + \left(  \rho \, \theta_{0,2} U^\beta U^\gamma +  \frac{1}{3}  \, \rho \,  c^2 \, \theta_{1,2} h^{\beta \gamma} \right) \tilde{\lambda}_1 \right]   
 \, .
 \end{split}
\end{align}
So for the Maxwellian Iteration we can repeat the passages of section 5 with the new expression \eqref{9a} for $I^{\beta \gamma}$. In particular, $\partial_\alpha A^{\alpha \beta \gamma}_E = I^{\beta \gamma}$, thanks to  $T^{\alpha \beta} - T^{\alpha \beta}_E= \Pi \,  h^{\alpha \beta} \, + \, t^{< \alpha \beta >_3}$, is an equation which  contracted with $h_\beta^{< \theta} \, h_\gamma^{\psi >}$, $h_\beta^{\theta} \, U_\gamma$,  $U_\beta U_\gamma$, $h_{\beta \gamma}$   gives respectively
\begin{align}\label{13a} 
\begin{split}
& h_\beta^{< \theta} \, h_\gamma^{\psi >} \, \partial_\alpha A^{\alpha \beta \gamma}_E =  - \,  \frac{1}{m \, \tau} \,  h_\beta^{< \theta} \, h_\gamma^{\psi >} \, \left( T^{\beta \gamma} - T^{\beta \gamma}_E \right)^{(OT)} = - \,  \frac{1}{m \, \tau} \, t^{(OT) < \theta \psi >_3}    \, , \\
& h_\beta^{\theta} \, U_\gamma \, \partial_\alpha A^{\alpha \beta \gamma}_E =  - h_\beta^{\theta} \, U_\gamma \, \frac{1}{m \, \tau} \left( T^{\beta \gamma} -  \, T^{\beta \gamma}_E \right)^{(OT)} \, - \, \frac{\rho \, c^4 \theta_{1,2}}{3 m \, \tau \, p} \, q^{(OT) \theta} = - \, \frac{\rho \, c^4 \theta_{1,2}}{3 m \, \tau \, p} \, q^{(OT) \theta}  \,  , \\
& \frac{U_\beta \, U_\gamma}{c^4} \, \partial_\alpha A^{\alpha \beta \gamma}_E =  \frac{1}{k_B \tau}  \left( \frac{e}{c^2} \, \tilde{\lambda}  \, + \,  \rho \, \theta_{0,2}  \tilde{\lambda}_1 \right)  \, , \\
& h_{\beta \gamma} \,   \partial_\alpha A^{\alpha \beta \gamma}_E  = - \, \frac{3 \, \Pi}{m \, \tau}  \, + \frac{1}{k_B \tau}  \left( 3 \, p \, \tilde{\lambda}  \, + \,  \rho \, c^2  \theta_{1,2}  \tilde{\lambda}_1 \right)   \, .
\end{split}
\end{align}
In the last two of these we can substitute $\tilde{\lambda}$, $\tilde{\lambda}_1$ from \eqref{8} in terms of $A-A_E$. After that, the first one of these new equations give the first interate of $A-A_E$; by substituting it in the second one, we obtain 
\begin{align}\label{8aa}
\frac{3 \, \Pi}{m \, \tau} = - \, h_{\beta \gamma} \,   \partial_\alpha A^{\alpha \beta \gamma}_E \, + \, \frac{ \frac{3 \, e \, p}{\rho^2 c^4 } \, - \, \theta_{1,2}}{\frac{e^2}{\rho^2 c^4} \, - \, \theta_{0,2} }  \, \frac{U_\beta \, U_\gamma}{c^6} \, \partial_\alpha A^{\alpha \beta \gamma}_E \, . 
\end{align}
Finally, \eqref{13a}$_{1,2}$ and \eqref{8aa} can be used in the usual way to desume 
\begin{align}\label{15a} 
\begin{split}
& \Pi^{(OT) } = - \, \nu \, \partial_\alpha \, U^\alpha \, , \, q^{(OT) \theta} = - \, \chi \, h^{\theta \mu} \, \left( \partial_\mu \, T \, - \, \frac{T}{c^2} \, U^\gamma \, \partial_\gamma \, U_\mu \right) \, , \\ 
& t_{(OT)  < \beta \gamma >}= 2 \, \mu  \, \partial_{< \beta} U_{\gamma >} \, ,
\end{split}
\end{align}
with explicit expressions of  the bulk viscosity $\nu$, the heat conductivity $\chi$ and the shear viscosity $\mu$.  They don't depend on the number $N$ of the model from which they were obtained. It is not necessary to compare them with those of the Eckart approach,  because the variables in the Eckart approach are different from those in the  Landau-Lifshiz description. In conclusion, we can say that Ordinary Thermodynamics in the  Landau-Lifshiz description has the balance equations
\begin{align}\label{52}
\begin{split}
& \partial_\alpha \left[ \rho \, U^\alpha \,  - \, \chi \, h^{\theta \mu} \, \left( \partial_\mu \, T \, - \, \frac{T}{c^2} \, U^\gamma \, \partial_\gamma \, U_\mu \right) \right]=0 \, , \\
& \partial_\alpha \left[ T^{\alpha \beta}_E \,   - \, \nu \, h^{\alpha \beta} \, \partial_\mu \, U^\mu \, \, + \,  2 \, \mu  \, h^{\alpha \mu} h^{\beta \nu} \partial_{< \mu} U_{\nu >_3}\right]=0 \, , 
\end{split}
\end{align}
with $\chi$, $\nu$ and $\mu$ obtained above.

\appendix

\section{The passage from $J_{m,n}$ to the modified Bessel Functions}

The use of the modified Bessel Functions is important because programs as "Mathematica" can easily deal with them. For the passage from $J_{m,n}$ to them, we have the folowing \\
{\bf Theorem 1:} "We have the following transformation equations: 
 \begin{align}\label{43}
\begin{split}
& K_n (\gamma) = \sum_{r=0}^{\left[ \frac{n}{2} \right]} \left( \begin{matrix}
n \\2r
\end{matrix}\right) J_{2r,n-2r} (\gamma)   \quad \mbox{and vice versa} \\
& J_{0,n} = \frac{1}{2^{n-1}} \sum_{r=0}^{\left[ \frac{n-1}{2} \right]}  \left( \begin{matrix}
n \\ r
\end{matrix}\right)  K_{n-2r} (\gamma) + \frac{1}{2^{n}}  \left( \begin{matrix}
n \\ \frac{n}{2}
\end{matrix}\right)   K_0 (\gamma) \, , 
 \end{split}
\end{align}
where the last term is present only if $n$ is even. The other $J_{m,n}$ can be easily obtained from \eqref{43} thanks to the recurrence relation 
 \begin{align}\label{44}
 J_{m+2,n} = J_{m,n+2} - J_{m,n} \, " . 
\end{align}
This last equation comes by substituting $\sinh^2 s$ with $\cosh^2 s - 1$ in the definition of $J_{m,n}$. \\
{\bf Proof:}  We have
 \begin{align*}
\begin{split}
& e^s = \cosh \, s + \sinh \, s \, , \, e^{-s} = \cosh \, s - \sinh \, s \quad \rightarrow \\
& \cosh \, (ns) = \frac{e^{ns} + e^{- ns}}{2} = \frac{1}{2} \, \left[ \left( \cosh \, s + \sinh \, s \right)^n + \left( \cosh \, s +- \sinh \, s \right)^n \right] = \\
& =  \frac{1}{2} \sum_{p=0}^{n}  \left( \begin{matrix}
n \\ p
\end{matrix}\right) \left[ \sinh^p s \, \cosh^{n-p} s + (-1)^p \sinh^p s \, \cosh^{n-p} s \right] \, .
 \end{split}
\end{align*}
Here only the terms with $p$ even doesn't simplify each other; By putting $p=2r$ we obtain \eqref{43}$_1$. \\
From its definition, we have 
 \begin{align*}
\begin{split}
\cosh \, s = \frac{e^{s} + e^{- s}}{2} \quad \rightarrow \quad  cosh^n \, s =  \frac{1}{2^n} \, \sum_{p=0}^{n}  \left( \begin{matrix}
n \\ p
\end{matrix}\right) e^{-ps} e^{(n-p)s} \, .
 \end{split}
\end{align*}
If $n$ is odd, this can be written as 
 \begin{align*}
\begin{split}
& cosh^n \, s =  \frac{1}{2^n}  \left[ \sum_{p=0}^{\frac{n-1}{2}}  \left( \begin{matrix}
n \\ p
\end{matrix}\right) e^{(n-2p)s} +  \sum_{p=\frac{n+1}{2}}^{n}  \left( \begin{matrix}
n \\ p
\end{matrix}\right)  e^{(n-2p)s} \right]\, .
\end{split}
\end{align*}
In the second summation we can change index according to the law $p=n-r$; so we obtain 
 \begin{align*}
\begin{split}
& cosh^n \, s =  \frac{1}{2^n} \left[  \sum_{p=0}^{\frac{n-1}{2}}  \left( \begin{matrix}
n \\ p
\end{matrix}\right)  e^{(n-2p)s} +  \sum_{r=0}^{\frac{n-1}{2}}  \left( \begin{matrix}
n \\ n-r
\end{matrix}\right) e^{- (n-2r)s} \right] \, .
\end{split}
\end{align*}
This proves \eqref{43}$_2$ if $n$ is odd. Instead of this, if $n$ is even, the above relation can be written as 
 \begin{align*}
\begin{split}
& cosh^n \, s =  \frac{1}{2^n}  \left[ \sum_{p=0}^{\frac{n-2}{2}}  \left( \begin{matrix}
n \\ p
\end{matrix}\right) e^{(n-2p)s} +  \left( \begin{matrix}
n \\ \frac{n}{2}
\end{matrix}\right)  + \sum_{p=\frac{n+2}{2}}^{n}  \left( \begin{matrix}
n \\ p
\end{matrix}\right)  e^{(n-2p)s} \right] \, .
\end{split}
\end{align*}
By proceeding as in the case with $n$ odd, we obtain \eqref{43}$_2$ with $n$ even. \\
In the sequel we will use the identity known in literature
 \begin{align}\label{45}
\gamma \, J_{m,n} = (n+m-1) \, J_{m-2,n+1} - n \, J_{m-2,n-1} \,  . 
\end{align}
The proof comes from in integration by parts (In the case $m \geq 2$)
 \begin{align*}
 \begin{split}
& J_{m,n} = \frac{-1}{\gamma} \int_{0}^{+ \infty} \left( e^{- \gamma \, \cosh \, s} \right)' \,  \sinh^{m-1} s \, \cosh^n \, s \, d \, s = =  \frac{-1}{\gamma}  \left| e^{- \gamma \, \cosh \, s}  \,  \sinh^{m-1} s \, \cosh^n \, s \right|_0^{+ \infty} + \\
& + \frac{1}{\gamma}  \int_{0}^{+ \infty} e^{- \gamma \, \cosh \, s}  \,  \left[  (m-1) \, \sinh^{m-2} s \, \cosh^{n+1} \, s + n \, \sinh^m s \,   \cosh^{n-1} \, s  \right] \, d \, s = \\
& = \frac{1}{\gamma}  \left[ (m-1) \, J_{m-2,n+1} \, + \, n \, J_{m,n-1} \right]  . 
\end{split}
\end{align*}
By using \eqref{44} with $m-2$ instead of $m$ and $n-1$ instead of $n$, we obtain \eqref{45}. \\
Let us consider now \eqref{45} with $m=2$ and substitute in it $J_{2,n}$ from \eqref{44} with $m=0$; we obtain a relation from which we desume 
 \begin{align}\label{46}
  J_{0,n+2}   = \frac{n+1}{\gamma} \, J_{0,n+1} + J_{0,n}   - \frac{n}{\gamma} \, J_{0,n-1} \quad \rightarrow \quad  J_{0,2}   =\frac{1}{\gamma} \, J_{0,1}  +  J_{0,0}   \,  . 
\end{align}
This proves the following \\
{\bf Theorem 2:} "The functions $J_{0,0}$ and $J_{0,1}$ are arbitrary, while $J_{0,2}$ is expressed as a linear combination of them trough \eqref{46}$_2$; all the other $J_{0,n}$ are expressed through \eqref{46}$_1$ as a linear combination of the three terms preceding it". \\
Let us define now 
 \begin{align}\label{47}
G^* = \frac{J_{0,0}}{J_{0,1}}  = \frac{K_0}{K_1} \quad \rightarrow \quad \frac{d \, G^*}{d \, \gamma} = - 1  + \, \frac{G^*}{\gamma} \, + \left(G^* \right)^2 \,  . 
\end{align}
To prove the second part of this equations, we calculate 
 \begin{align*}
\frac{d \, }{d \, \gamma} \,  \frac{J_{0,0}}{J_{0,1}}  = \frac{\frac{d \, J_{0,0}}{d \, \gamma}  }{J_{0,1}}  - \, \frac{J_{0,0} \, \frac{d \, J_{0,1}}{d \, \gamma}}{ \left( J_{0,1} \right)^2} = - 1  + \, G^* \, \frac{J_{0,2}}{J_{0,1}} \,  . 
\end{align*}
By using \eqref{46}$_2$, we obtain \eqref{47}$_2$. So the derivative of $G^*$ with respect to $\gamma$ is a function of the same $G^*$. Now in literature it is commonly used the function $G=\frac{K_3}{K_2}$; so we have to find the relation between $G$ and $G^*$. From \eqref{43}$_1$ we have 
 \begin{align}\label{48}
G = \frac{ J_{0,3}  + 3 J_{2,1} }{ J_{0,2}  +  J_{2,0}  } =\frac{4 \, J_{0,3}  - 3 J_{0,1} }{ 2 \, J_{0,2}  -  J_{0,0}  }  =  \frac{  \frac{4}{\gamma} \, J_{0,0}   + \left( 1 + \frac{8}{\gamma^2} \right) J_{0,1} }{ \frac{2}{\gamma} \, J_{0,1}     +  J_{0,0}  }  = \frac{  \frac{4}{\gamma} \, G^*   +  1 + \frac{8}{\gamma^2}  }{G^* + \frac{2}{\gamma}   }  \, , 
\end{align}
where in the first passage we have used \eqref{44} and in the second passage \eqref{46}. The inverse of this relation is 
\begin{align}\label{49}
G^* = - \frac{2}{\gamma} + \frac{\gamma}{\gamma G -4} \, . 
\end{align}

\section{The meaning of the traceless part of a symmetric tensor of order n}
Let us consider the number $a_{r,n}= \frac{n+1-2 \, r}{4^r (n+1)} \left( \begin{matrix}
n+1 \\
r
\end{matrix}\right)$ and the decomposition 
\begin{align}\label{39}
T^{\alpha_1 \cdots \alpha_n} = \sum_{r=0}^{\left[ \frac{n}{2}\right]} a_{r,n} 
g^{( \alpha_1 \alpha_2} \cdots  g^{\alpha_{2r-1} \alpha_{2r}} T^{< \alpha_{2r+1} \cdots \alpha_n > ) \mu_1 \mu_2 \cdots \mu_{2r-1} \mu_{2r}} g_{\mu_1 \mu_2} \cdots g_{\mu_{2r-1} \mu_{2r}} \, . 
\end{align}
In particular, we have 
\begin{align*}
& a_{0,n}= 1 \quad \forall \, n \, , \, a_{1,2}= \frac{1}{4} \, , \, a_{1,3}= \frac{1}{2} \quad \rightarrow \quad T^{\alpha_1  \alpha_2} = T^{< \alpha_1  \alpha_2 >} + \frac{1}{4} \, g^{\alpha_1  \alpha_2} T^\mu_\mu \, ,   \\  
& T^{\alpha_1  \alpha_2 \alpha_3} = T^{< \alpha_1  \alpha_2 \alpha_3 >} + \frac{1}{2} \, g^{( \alpha_1  \alpha_2} T^{\alpha_3 ) \mu \nu} g_{\mu \nu} \quad \rightarrow  \\
&  T^{< \alpha_1  \alpha_2  >} g_{\alpha_1 \alpha_2} = 0 \, , \,
T^{< \alpha_1  \alpha_2 \alpha_3 >} g_{\alpha_2 \alpha_3} = 0 \, . 
\end{align*}
Therefore, $T^{< \alpha_1  \alpha_2 >}$ and $T^{< \alpha_1  \alpha_2 \alpha_3 >}$ are the simmetric traceless parts of $T^{\alpha_1  \alpha_2}$ and $T^{ \alpha_1  \alpha_2 \alpha_3 }$ respectively. Let us suppose with an iterative procedure that this property holds up to a given number $n=$, i.e., that $T^{< \alpha_{2r+1} \cdots \alpha_n > \mu_1 \mu_2 \cdots \mu_{2r-1} \mu_{2r}} g_{\mu_1 \mu_2} \cdots g_{\mu_{2r-1} \mu_{2r}}$, appearing in \eqref{39}, are the simmetric traceless parts of \\
$T^{ \alpha_{2r+1} \cdots \alpha_n  \mu_1 \mu_2 \cdots \mu_{2r-1} \mu_{2r}} g_{\mu_1 \mu_2} \cdots g_{\mu_{2r-1} \mu_{2r}}$ and let us prove that this is true also with $n+2$ instead of $n$. So let us write \eqref{39}  with $n+2$ instead of $n$
\begin{align*}
T^{\alpha_1 \cdots \alpha_{n+2}} = \sum_{r=0}^{\left[ \frac{n+2}{2}\right]} a_{r,n+2} 
g^{( \alpha_1 \alpha_2} \cdots  g^{\alpha_{2r-1} \alpha_{2r}} T^{< \alpha_{2r+1} \cdots \alpha_{n+2}> ) \mu_1 \mu_2 \cdots \mu_{2r-1} \mu_{2r}} g_{\mu_1 \mu_2} \cdots g_{\mu_{2r-1} \mu_{2r}} \, , 
\end{align*}
and let us consider it the definition of $T^{< \alpha_1 \cdots \alpha_{n+2}>}$ (it appears in the term with $r=0$). By contracting this relation with $g_{\alpha_{n+1} \alpha_{n+2}}$ we obtain 
\begin{align*}
& T^{\alpha_1 \cdots \alpha_{n+2}} g_{\alpha_{n+1} \alpha_{n+2}} = T^{< \alpha_1 \cdots \alpha_{n+2} >} g_{\alpha_{n+1} \alpha_{n+2}} + \\
& + g_{\alpha_{n+1} \alpha_{n+2}} \sum_{r=1}^{\left[ \frac{n+2}{2}\right]} a_{r,n+2} 
g^{( \alpha_1 \alpha_2} \cdots  g^{\alpha_{2r-1} \alpha_{2r}} T^{< \alpha_{2r+1} \cdots \alpha_{n+2}> ) \mu_1 \mu_2 \cdots \mu_{2r-1} \mu_{2r}} g_{\mu_1 \mu_2} \cdots g_{\mu_{2r-1} \mu_{2r}} = \\
& = T^{< \alpha_1 \cdots \alpha_{n+2} >} g_{\alpha_{n+1} \alpha_{n+2}} +  \sum_{r=1}^{\left[ \frac{n+2}{2}\right]} \frac{a_{r,n+2}}{(n+1)(n+2)} 
\left[ g_{\alpha_{n+1} \alpha_{n+2}} 2r  g^{\alpha_{n+1} \alpha_{n+2}}  g^{( \alpha_1 \alpha_2} \cdots  g^{\alpha_{2r-3} \alpha_{2r-2}} \cdot \right. \\
& T^{< \alpha_{2r-1} \cdots \alpha_{n}> ) \mu_1 \mu_2 \cdots \mu_{2r-1} \mu_{2r}} g_{\mu_1 \mu_2} \cdots g_{\mu_{2r-1} \mu_{2r}}  + \\
& + \left. 2r (2r-2) g^{\alpha_{n+1} \alpha_{n+2}}  g_{\alpha_{n+1}}^{( \alpha_1} \cdots  g_{\alpha_{n+2}}^{\alpha_{2r-2}} T^{< \alpha_{2r-1} \cdots \alpha_{n}> ) \mu_1 \mu_2 \cdots \mu_{2r-1} \mu_{2r}} g_{\mu_1 \mu_2} \cdots g_{\mu_{2r-1} \mu_{2r}} \right] + \\
& + 4r (n+2 - 2r) g_{\alpha_{n+1} \alpha_{n+2}}  g^{\alpha_{n+2} ( \alpha_1}  \cdots  g^{\alpha_{2r-2} \alpha_{2r-1}} T^{< \alpha_{2r} \cdots \alpha_{n} ) \alpha_{n+1}> \mu_1 \mu_2 \cdots \mu_{2r-1} \mu_{2r}} \cdot \\
& \left. \cdot  g_{\mu_1 \mu_2} \cdots g_{\mu_{2r-1} \mu_{2r}} \right] = T^{< \alpha_1 \cdots \alpha_{n+2} >} g_{\alpha_{n+1} \alpha_{n+2}} + \\
& +  \sum_{r=1}^{\left[ \frac{n+2}{2}\right]} \frac{a_{r,n+2}}{(n+1) (n+2)} 
g^{( \alpha_1 \alpha_2} \cdots  g^{\alpha_{2r-3} \alpha_{2r-2}}  
T^{< \alpha_{2r-1} \cdots \alpha_{n}> ) \mu_1 \mu_2 \cdots \mu_{2r-1} \mu_{2r}} g_{\mu_1 \mu_2} \cdots g_{\mu_{2r-1} \mu_{2r}}  \cdot \\
& \cdot 4r (n+1 )(n+3-r)  =  T^{< \alpha_1 \cdots \alpha_{n+2} >} g_{\alpha_{n+1} \alpha_{n+2}} + \\
& + \sum_{R=0}^{\left[ \frac{n}{2}\right]} \frac{a_{R+1,n+2}}{(n+1)(n+2)} 4(R+1) (n+2-R) \,
g^{( \alpha_1 \alpha_2} \cdots  g^{\alpha_{2R-1} \alpha_{2R}}  
T^{< \alpha_{2R+1} \cdots \alpha_{n}> ) \mu_1 \mu_2 \cdots \mu_{2R-1} \mu_{2R} \alpha_{n+1} \alpha_{n+2} }   \cdot \\
& \cdot g_{\mu_1 \mu_2} \cdots g_{\mu_{2R-1} \mu_{2R}} g_{\alpha_{n+1} \alpha_{n+2}} 
\, ,
\end{align*}
where in the last passage we have changed index according to the law $r=1+R$. 
Since $a_{R+1,n+2} \frac{ 4(R+1) (n+2-R)}{(n+1)(n+2)} =  a_{R,n}$, the left hand side and the last term in the right hand side simplify each other and there remains  
$ T^{< \alpha_1 \cdots \alpha_{n+2} >} g_{\alpha_{n+1} \alpha_{n+2}}=0$. This proves that $ T^{< \alpha_1 \cdots \alpha_{n+2} >}$ defined as above is the simmetric traceless part of $ T^{\alpha_1 \cdots \alpha_{n+2}}$. Thanks to this fact, \eqref{3}$_1$ for monoatomic gases can be substituted by
\begin{align*}
\partial_\alpha \, A^{< \alpha \alpha_1 \cdots \alpha_n >} = I^{< \alpha_1 \cdots \alpha_n >} \,  .
\end{align*}
(We have simply added to the equation with the index $n$ a linear combination of those with a less value of $n$).

\end{document}